\newcommand*{\ATLASLATEXPATH}{latex/}
\pdfoutput=1
\documentclass[cernpreprint,texlive=2011,txfonts,UKenglish]{\ATLASLATEXPATH atlasdoc}
\usepackage[subfigure=true,block=space,biblatex=true]{\ATLASLATEXPATH
atlaspackage}
\usepackage{\ATLASLATEXPATH atlasbiblatex}

\usepackage{\ATLASLATEXPATH atlascover}

\usepackage{\ATLASLATEXPATH atlascontribute}

\usepackage[unit=false]{\ATLASLATEXPATH atlasphysics}

\usepackage{epstopdf}

\graphicspath{{logos/}{images/}}

\usepackage{topq_2015_15-defs}
\AtlasTitle{Measurement of the inclusive cross-sections of single top-quark and 
top-antiquark $t$-channel production in $pp$ collisions at $\sqrt{s}$ = \SI{13}{\TeV} with the ATLAS detector}

\author{The ATLAS Collaboration}

\AtlasRefCode{TOPQ-2015-15}

\AtlasNote{ATL-COM-PHYS-2016-271}

\PreprintIdNumber{CERN-EP-2016-197}

\AtlasJournal{JHEP}
\AtlasJournalRef{JHEP 04 (2017) 086}
\AtlasDOI{10.1007/JHEP04(2017)086}

\AtlasAbstract{%
A measurement of the $t$-channel single-top-quark and single-top-antiquark production cross-sections in the lepton+jets
channel is presented, using \SI{3.2}{\per\fb} of proton--proton collision data at a centre-of-mass energy of \SI{13}{\TeV}, recorded
with the ATLAS detector at the LHC in 2015.
Events are selected by requiring one charged lepton (electron or muon), missing transverse momentum, 
and two jets with high transverse momentum, exactly one of which is required to be $b$-tagged.
Using a binned maximum-likelihood fit to the discriminant distribution of a neural network,
the cross-sections are determined to be
$\sigma(tq)      =  156 \pm 5  \, (\mathrm{stat.}) \pm 27 \, (\mathrm{syst.}) \pm 3\,(\mathrm{lumi.})~\text{pb}$
for single top-quark production and 
$\sigma(\tbar q) =   91 \pm 4  \, (\mathrm{stat.}) \pm 18 \, (\mathrm{syst.}) \pm 2\,(\mathrm{lumi.})~\text{pb}$
for single top-antiquark production, assuming a top-quark mass of \SI{172.5}{\GeV}.
The cross-section ratio is measured to be $R_t = \sigma(tq)/\sigma(\tbar q) = 1.72 \pm 0.09  \, (\mathrm{stat.}) \pm
0.18 \, (\mathrm{syst.})$. All results are in agreement with Standard Model predictions.
}

\AtlasCoverSupportingNote{INT note}{https://cds.cern.ch/record/2129154}
\AtlasCoverCommentsDeadline{August 10, 2016}

\AtlasCoverAnalysisTeam{Dominic Hirschb\"uhl, Gunnar J\"akel, Phillipp Tepel, Wolfgang Wagner}

\AtlasCoverEdBoardMember{Matthias Danninger}
\AtlasCoverEdBoardMember{Pierre Savard~(chair)}
\AtlasCoverEdBoardMember{Timothee Theveneaux-Pelzer}

\AtlasCoverEgroupEditors{atlas-topq-2015-15-editors@cern.ch}

\AtlasCoverEgroupEdBoard{atlas-topq-2015-15-editorial-board@cern.ch}

\begin{document}
\hypersetup{pageanchor=false}
\maketitle
\hypersetup{pageanchor=true}
\clearpage
\section{Introduction}
\label{sec:intro}
After its restart in 2015, the Large Hadron Collider (LHC)~\cite{Evans:2008zzb}
has been producing proton--proton ($pp$) collisions at a centre-of-mass energy
of $\sqrt{s}=\SI{13}{\TeV}$, giving the collider experiments access to a so far 
unexplored kinematic range. 
It is important to measure all accessible Standard Model (SM) processes 
at the new centre-of-mass energy, compare the results to the corresponding theoretical 
SM predictions, and look for deviations which might result from energy-dependent
non-SM couplings.
In this article, inclusive cross-section measurements of the dominant 
single-top-quark production mechanism are presented.

\begin{figure}[!h!tpb]
  \centering
\subfigure[]{
  \includegraphics[width=0.21\textwidth]{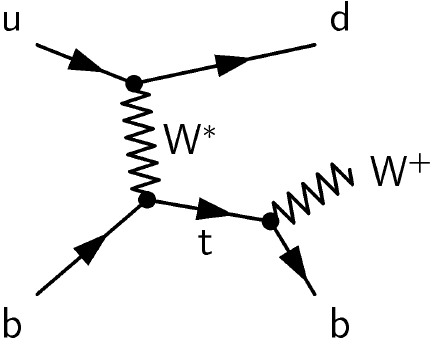}
  \label{subfig:top_quark}
}
\hspace*{0.03\textwidth}
\subfigure[]{
  \includegraphics[width=0.21\textwidth]{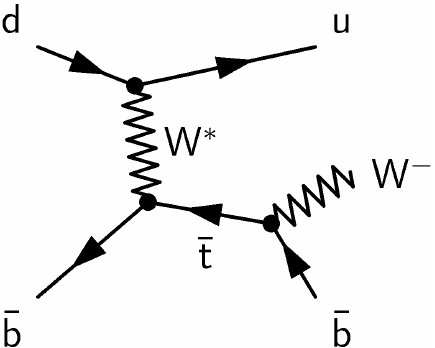}
  \label{subfig:top_antiquark}
}
\caption{Representative leading-order Feynman diagrams of \subref{subfig:top_quark} 
   single-top-quark production and \subref{subfig:top_antiquark} single-top-antiquark 
   production via the $t$-channel exchange of a virtual $W$ boson ($W^*$), including the decay of the top quark 
   and top antiquark, respectively.}
\label{fig:Feynman_tchan}
\end{figure}

At leading order (LO) in perturbation theory, single top-quark production is described by three 
subprocesses that are distinguished by the virtuality of the exchanged $W$ boson. 
The dominant process is the $t$-channel exchange depicted in Figure~\ref{fig:Feynman_tchan}, 
which is the subject of the measurements presented in this article.
A light quark from one of the colliding protons interacts with a $b$-quark from 
another proton by exchanging a virtual $W$ boson. Since the valence $u$-quark density of the proton 
is about twice as high as the valence $d$-quark density,
the production cross-section of single top-quarks $\sigma(tq)$ 
is expected to be higher than the cross-section of top-antiquark production
$\sigma(\bar{t}q)$.
At LO, the subleading single-top-quark processes are the associated production of a $W$ boson 
and a top quark~($Wt$) and the $s$-channel production of $t\bar{b}$ and $\bar{t} b$.

In this article, measurements of $\sigma(tq)$ and $\sigma(\bar{t}q)$ in
proton--proton collisions at a centre-of-mass energy of $\sqrt{s} =
\SI{13}{\TeV}$ are presented.
The analysis is based on the ATLAS data set collected in 2015 corresponding
to an integrated luminosity of \SI{3.2}{\per\fb}. Separate measurements of 
$tq$ and $\bar{t}q$ production provide sensitivity to the parton distribution 
functions (PDFs) of the $u$-quark and the $d$-quark~\cite{Alekhin:2015cza}, exploiting the different 
initial states of the two processes as shown in Figure~\ref{fig:Feynman_tchan}.
In addition, the cross-section ratio $R_t \equiv  \sigma(tq)/\sigma(\bar{t}q)$
is measured, featuring smaller systematic uncertainties than the
individual cross-sections because of partial cancellations of common
uncertainties.

In general, measurements of single top-quark production provide insights into the properties 
of the $Wtb$ vertex. The cross-sections are proportional to the square of the coupling at the 
production vertex. In the SM, the coupling is given by the Cabibbo--Kobayashi--Maskawa (CKM) 
matrix element $V_{tb}$~\cite{Cabibbo:1963yz,Kobayashi:1973fv} multiplied by 
the universal electroweak coupling constant.
Non-SM contributions can be encapsulated by an additional left-handed form factor
\fl~\cite{AguilarSaavedra:2008zc}, assumed to be real.
The sensitivity for these non-SM contributions could be increased for the
higher centre-of-mass energy, if there is new physics at high scales.
The combined cross-section $\sigma(tq+\bar{t}q)$ is determined as the sum of
$\sigma(tq)$ and $\sigma(\bar{t}q)$ and used to determine $\fl \cdot |V_{tb}|$.
All measurements presented in this paper are based on the assumption
that the production and the decay of top quarks via $Wts$ and $Wtd$ 
vertices is suppressed due to the fact that the CKM matrix elements 
$V_{ts}$ and $V_{td}$ are much smaller than $V_{tb}$.
Currently, the most precise determination of $\fl \cdot |V_{tb}|$ has an uncertainty 
of \SI{4}{\%}, obtained from a combination of measurements performed by the CMS
Collaboration~\cite{CMS-TOP-12-038} under the same assumption as the one stated above.

In $pp$ collisions at $\sqrt{s}=\SI{13}{\TeV}$, the predicted production cross-section of
the $t$-channel single-top-quark process 
is $\sigma(tq)= 136.0^{+5.4}_{-4.6}\;$pb for top-quark production and
$\sigma(\bar{t}q)=81.0^{+4.1}_{-3.6}\;$pb for top-antiquark production. 
These predictions have been calculated for a top-quark mass of \SI{172.5}{\GeV} at next-to-leading order
(NLO)~\cite{Campbell:2009ss} in perturbative QCD using Hathor v2.1~\cite{Kant:2014oha}.
The uncertainties connected with PDFs and the strong coupling constant, \alphas,
are calculated using the PDF4LHC prescription~\cite{Botje:2011sn} with the
\mstw~NLO~\cite{Martin:2009iq,Martin:2009bu}, \ct~NLO~\cite{Lai:2010vv} and \nnpdftwo~NLO~\cite{Ball:2012cx} PDF sets,
and are added in quadrature to the scale uncertainty.
The cross-sections of all three single-top-quark production processes
have also been calculated at approximate next-to-next-to-leading-order (NNLO) precision
~\cite{Kidonakis:2011wy,Kidonakis:2010ux,Kidonakis:2010tc}.
NNLO results are available for the $t$-channel cross-section~\cite{Brucherseifer:2014ama}
and the NNLO/NLO $K$-factor is 0.985. However, the NLO calculation of this process  
features a more comprehensive uncertainty treatment,
including a complete treatment of the PDF and scale uncertainties, 
compared to the NNLO one, and is therefore used to extract $\fl \cdot |V_{tb}|$. 
 
In this analysis, the event selection targets $tq$ and $\tbar q$ events
with leptonically decaying $W$ bosons. The lepton is either an electron or a muon, 
where events involving $W\rightarrow \tau\nu$ decays with a subsequent
decay of the $\tau$ lepton to $e\nu_e\nu_\tau$ or $\mu\nu_\mu\nu_\tau$ are included in the signal.
The experimental signature of selected events is thus given by one prompt isolated electron or muon, missing
transverse momentum, \MET, and two hadronic jets with high transverse momentum,
\pT,\footnote{ATLAS uses a right-handed coordinate system with its origin at the nominal interaction point in the
centre of the detector and the $z$-axis is along the beam direction;
the $x$-axis points towards the centre of the LHC ring and 
the $y$-axis points upwards. The pseudorapidity 
$\eta$ is defined as $\eta=-\ln[\tan(\theta/2)]$, where the polar angle 
$\theta$ is measured with respect to the $z$-axis. The azimuthal angle, 
$\phi$, is measured with respect to the $x$-axis.
Transverse momentum and energy are defined as $\pT = 
p\sin\theta$ and $\ET = E\sin\theta$, respectively. The $\Delta R$ distance 
in ($\eta$,$\phi$) space is defined as $\Delta R=\sqrt{(\Delta\eta)^2+(\Delta\phi)^2}$.}
where one of these jets originates from a $b$-quark ($b$-jet) and the second one
is produced primarily in the forward direction.
The presence of additional jets is vetoed to suppress background from 
\ttbar production.

Several other processes feature the same signature as single-top-quark events; the main backgrounds
are $W+$jets production and top-quark--antiquark (\ttbar) pair production.
In order to improve the sensitivity of the signal extraction, 
an artificial neural network (NN)~\cite{feindt-2004} is used to discriminate
between signal and background events, following the same strategy that was used 
in comprehensive measurements of $t$-channel single top-quark production at $\sqrt{s}=$\SI{7}{\TeV}~\cite{TOPQ-2012-21}. 

This article is organised as follows. 
Section~\ref{sec:samples} gives an overview
of the data and simulated event samples
that are used in the analysis.
The definitions of physics objects are given in Section~\ref{sec:obj} 
and the event selection criteria as well as the definition of the signal 
and validation regions are presented in Section~\ref{sec:sel}. 
Section~\ref{sec:background} describes
the estimation of the background processes 
and compares the predicted kinematic distributions to data.
Section~\ref{sec:nn} discusses the discriminating variables and 
the training and the performance of the NN used to improve the measurement sensitivity,
while in Section~\ref{sec:systematics} the estimation of systematic uncertainties is 
discussed.
Section~\ref{sec:stat} is devoted to the 
statistical analysis and Section~\ref{sec:result} to the measurement of the signal cross-sections and 
their ratio, and the  
extraction of $\fl \cdot |V_{tb}|$. 
Finally, the conclusion is given in Section~\ref{sec:conclusion}.

\section{Data and simulation samples}
\label{sec:samples}
The ATLAS experiment~\cite{PERF-2007-01} at the LHC is a multi-purpose particle detector
with a forward-backward symmetric cylindrical geometry and a near $4\pi$ coverage in 
solid angle.
It consists of an inner tracking detector surrounded by a thin superconducting solenoid
providing a \SI{2}{\tesla} axial magnetic field, electromagnetic and hadron calorimeters, and a muon spectrometer.
The inner tracking detector (ID) covers the pseudorapidity range $|\eta| < 2.5$.
It consists of silicon pixel, silicon microstrip, and transition-radiation tracking detectors.
The innermost layer of the pixel detector, the insertable B-layer~\cite{ATLAS-TDR-19}, 
was added between Run 1 (2009-2013) 
and Run 2 of the LHC at a radius of \SI{33}{\mm} around a new and thinner beam pipe.
Lead/liquid-argon (LAr) sampling calorimeters provide electromagnetic (EM) energy measurements
with high granularity.
A hadron (iron/scintillator-tile) calorimeter covers the central pseudorapidity range ($|\eta| < 1.7$).
The endcap and forward regions are instrumented with LAr calorimeters
for both the EM and hadronic energy measurements up to $|\eta| = 4.9$.
The muon spectrometer surrounds the calorimeters and is based on
three large air-core toroid superconducting magnets with eight coils each.
Its bending power ranges from \num{2.0} to \SI{7.5}{\tesla\metre}.
It includes a system of precision tracking chambers and fast detectors for triggering.
A two-level trigger system~\cite{ATL-DAQ-PUB-2016-001} is used to select events.
The first-level trigger is implemented in hardware and uses a subset of the detector information
to reduce the accepted rate to at most \SI{75}{\kilo\hertz}.
This is followed by a software-based high-level trigger (HLT),
which has access to full detector granularity and is used to further reduce the 
event rate to \SI{1}{\kilo\hertz}.

This analysis is performed using $pp$ collision data 
recorded at a centre-of-mass energy of $\sqrt{s}=\SI{13}{\TeV}$ with the ATLAS detector
in 2015 in the periods when the LHC was operating with 25 ns bunch spacing. 
Only the periods in which all the subdetectors were operational are considered, resulting in a
data sample with a total integrated luminosity of ${\cal L} = \SI{3.2}{\per\fb}$.  

All generated samples are passed through the simulation of the ATLAS 
detector~\cite{SOFT-2010-01} based on \GEANT{}4~\cite{Agostinelli:2002hh}.
The same offline reconstruction methods 
used with data events are applied to the simulated events samples. 
Minimum-bias events generated by \PYTHIA8~\cite{Sjostrand:2007gs} are used to 
simulate multiple $pp$ interactions in the same and nearby bunch crossings (pile-up).
The simulated pile-up events are reweighted to reproduce the luminosity spectrum in the data.

Electroweak $t$-channel single-top-quark production can be simulated in
different schemes concerning the treatment of the initial $b$-quark.
In the five-flavour scheme (5FS) the $b$-quarks are treated massless and 
the LO Feynman diagram is represented by the $2 \rightarrow 2$ process 
with a $b$ quark in the initial state, depicted in
Figure~\ref{fig:Feynman_tchan}. In the four-flavour scheme (4FS), 
the PDFs only contain parton distributions for the quarks lighter than the $b$-quark
and therefore the LO Feynman diagram is represented by a $2 \rightarrow 3$
process including the $g \rightarrow b\bar{b}$ splitting in the initial state.
In this scheme, the $b$-quarks are treated massive.

Signal $t$-channel single-top-quark events are generated in the 4FS using the 
\powheg V1 (r2556)~\cite{Frederix:2012dh, Alioli:2010xd} generator
Events are generated with the fixed four-flavour PDF 
set \ctf~\cite{Lai:2010vv} and the renormalisation and factorisation
scales, $\mu_{\mathrm r}$, and , $\mu_{\mathrm f}$, 
are set following the recommendation given in Ref.~\cite{Frederix:2012dh}.
Top quarks are decayed at LO using \MADSPIN to preserve all spin correlations.
The parton shower, hadronisation, and the underlying event are 
modelled using the \PYTHIA6 (v6.428)~\cite{Sjostrand:2006za} generator and 
the Perugia2012 set of tuned parameters (P2012 tune)~\cite{Skands:2010ak}.
In order to study effects of the choice of parton-shower model,
the same events are showered using \herwigpp(v.2.7.1)~\cite{Bahr:2008pv} and 
the energy-extrapolated underlying event set of tuned parameters (UE-EE-5 tune)~\cite{Seymour:2013qka}.
A second NLO generator capable of generating $t$-channel single-top-quark events in the 4FS is 
\MGMCatNLO~\cite{Alwall:2014hca} (v2.2.2).
Samples are generated using the \ctf PDF set and $\mu_{\mathrm r}$ and $\mu_{\mathrm f}$ are set to be the same as those
implemented in \powheg.
Again, the top quarks produced in the ME are decayed using \MADSPIN to preserve all spin correlations.
The events are showered using \herwigpp(v.2.7.1) and the UE-EE-5 tune.
For the generation of SM single top-quarks in the $Wt$ and the $s$-channel ($t\bar{b}+\bar{t}b$)
the \powheg V1 (r2819) generator~\cite{Re:2010bp,Alioli:2009je} with the \ct PDF set is used.
Samples of \ttbar events are generated with the \powheg V2~(r3026)~\cite{Frixione:2007nw} 
and the \ct PDF set. 
The $\hdamp$ parameter, which controls the \pT of the first additional emission beyond the Born configuration, 
where its main effect is to regulate the high-\pT emission against which the \ttbar system recoils,
is set to the top-quark mass. The parton shower, hadronisation, and the
underlying event are added using \PYTHIA6 and the P2012 tune.

All top-quark processes are generated assuming a top-quark mass of \SI{172.5}{\GeV}.
The top quark is set to decay exclusively to 
$t \to Wb$, and the \evtgen v1.2.0 program~\cite{Lange:2001uf}
is used to model bottom- and charm-hadron decays.

To model the $W$+jets and $Z$+jets background, 
the \SHERPA v2.2.0~\cite{Gleisberg:2008ta} generator is used. 
Matrix elements are calculated for up to two partons at NLO
and up to four partons at LO using the Comix~\cite{Gleisberg:2008fv} and OpenLoops~\cite{Cascioli:2011va} ME
generators and merged with the \SHERPA parton shower~\cite{Schumann:2007mg} using the ME+PS@NLO
prescription~\cite{Hoeche:2012yf}. 
The \nnpdfthree PDF set~\cite{Ball:2014uwa} is used in conjunction with dedicated parton-shower tuning
developed by the \SHERPA authors. 

Diboson events, denoted $VV$, are also simulated using the \SHERPA v2.1.1 generator. Matrix elements
contain all diagrams with four electroweak vertices. They are calculated for zero partons at NLO and up to 
three partons at LO using the same methodology as for $W$/$Z$+jets production.
The \ct PDF set is used in conjunction with dedicated parton-shower tuning developed
by the \SHERPA authors.

The only background for which no simulated events are used is the multijet background.
Multijet events may be selected if a jet is misidentified as an isolated lepton 
("fake" lepton) or if a non-prompt lepton from a hadronic decay appears to be isolated.
In the electron channel, the matrix method is used, while in the muon channel,
the so-called \enquote{anti-muon} method 
is employed to estimate the multijet background~\cite{ATLAS-CONF-2014-058}.
More details are given in Section~\ref{sec:background}.

\section{Object reconstruction}
\label{sec:obj}
In this section, the reconstruction and selection of electrons, muons, jets and \MET is described.

Electron candidates are defined as clusters of energy deposits in the electromagnetic calorimeter associated 
with a well-measured track fulfilling several quality requirements~\cite{PERF-2013-03,ATLAS-CONF-2014-032}.
They are required to satisfy $\pT > \SI{30}{\GeV}$ and 
$\mathrm{|\eta_{\mathrm clus}| <}$ 2.47, where $\mathrm{\eta_{\mathrm clus}}$ is the
pseudorapidity of the cluster of energy deposits in the calorimeter.
Electron candidates in the calorimeter barrel--endcap transition region 
$1.37 < |\mathrm{\eta_{\mathrm clus}}| < 1.52$ are excluded.
Isolation criteria are applied to reduce background events, in which 
a hadronic jet is misidentified as a prompt electron or electrons from the decay of heavy quarks.
The criteria are optimised such that by adjusting the isolation threshold the selection efficiency 
of the isolation criteria is uniform across $\eta$.
It increases from \SI{90}{\%} for $\pT = \SI{25}{\GeV}$ to
\SI{99}{\%} for $\pT = \SI{60}{\GeV}$.
The \pT of all tracks within a cone of size $\Delta R = \sqrt{(\Delta\eta)^2+(\Delta\phi)^2} = 0.3$ 
around the electron direction, excluding the track belonging to the electron candidate (track isolation), 
is restricted to be below a threshold depending on the electron \pT.
In addition, calorimeter isolation in a cone size of 0.2 around the electron is required~\cite{TOPQ-2015-09}.

Muon candidates are reconstructed by matching track segments or complete tracks in the muon 
spectrometer with inner detector tracks.
The candidates are required to have a transverse
momentum $\pT > \SI{25}{\GeV}$ and to be in the pseudorapidity region $|\eta|<2.5$. 
Additional requirements on the transverse  
impact parameter significance of $|d_0/\sigma_{d_0}| < 3$ and on the 
longitudinal impact parameter ($z_{0}$) of $|\Delta z_{0}\sin\theta| < \SI{0.5}{mm}$ are
imposed. 
Isolation criteria similar to those for electron candidates are imposed.

Jets are reconstructed using the anti-$k_{t}$ algorithm~\cite{Cacciari:2008gp}
with a radius parameter of 0.4.
They are calibrated using a combination of  
an energy- and $\eta$-dependent simulation-based scheme
and a scheme based on data~\cite{ATLAS-CONF-2015-037}.
Only jets with $\pt>\SI{30}{\GeV}$  and $|\eta|<3.5$ are accepted.
The rapidity range is determined using a $W$+jets-dominated validation region
and defined by requiring good agreement between simulated and measured data .

If any jet is close to an electron, with $\Delta R<0.2$, the closest jet is removed, as in these
cases the jet and the electron are very likely to correspond to the same physical object. 
Remaining electron candidates 
overlapping with jets within $\Delta R<0.4$ are subsequently rejected.
To reduce contributions from muons which stem from heavy-flavour decays inside a jet, 
muons are removed if they are separated from the nearest jet by $\Delta R<0.4$. 
However, jets with fewer than three tracks and separated from a muon by $\Delta R<0.4$
are removed to reduce fake jets from muons depositing a large fraction of their energy in the calorimeters.

To discriminate between jets from the hard-scatter process and those from pile-up,
a discriminant called the jet vertex tagger (JVT)~\cite{PERF-2014-03} 
is constructed from tracking and vertexing information using a two-dimensional likelihood method.
The JVT variable is required to be larger than $0.64$ for the jets with $\pt < \SI{50}{\GeV}$ 
and $|\eta| < 2.4$, corresponding to \SI{92}{\%} efficiency and \SI{2}{\%} misidentification rate.

In this analysis, a $b$-tagging algorithm based on boosted decision trees which is 
optimised to reject $c$-quark jets as well as light-quark jets is used.
The $b$- and $c$-tagging efficiencies, and the mistag rate for the taggers, are measured using the 
methods described in Refs.~\cite{ATLAS-CONF-2014-046,ATLAS-CONF-2014-004}. 
The $b$-tagging algorithm has an efficiency of about \SI{60}{\%} for $b$-jets in 
simulated $\ttbar$ events, while \SI{0.06}{\%} of light-quark jets and \SI{4.7}{\%} of $c$-quark jets are
mistagged as $b$-quark jets. The algorithm can only be applied to jets within the coverage
of the ID, i.e. $|\eta| < 2.5$.

The magnitude of the missing transverse momentum vector is defined as $\MET = | \vec{E}_{\mathrm{T}}^{\mathrm{miss}} |$, 
where $\vec{E}_{\mathrm{T}}^{\mathrm{miss}}$ is calculated using 
the calibrated three-dimensional calorimeter energy clusters associated with the selected jets 
together with either the calibrated calorimeter energy cluster associated with an electron 
or the $\pT$ of a muon track (hard components).
Contributions from soft particles, not associated with these identified particles, are
accounted for using tracks associated to the vertex but not associated with a
jet, electron, or muon (soft components).

\section{Event selection}
\label{sec:sel}

Events are considered only if they are accepted by at least one of two single-muon or
single-electron triggers~\cite{TRIG-2012-03}.
Events in the electron channel are triggered by a  
calorimeter cluster matched to a track, and the trigger electron object is
required to have either $\et > \SI{60}{\GeV}$ or $\et > \SI{24}{\GeV}$ and satisfy isolation criteria. 
Events in the muon channel are triggered by either requiring an isolated muon with $\pT >
\SI{20}{\GeV}$ or requiring a muon with $\pT > \SI{50}{\GeV}$.

Only events containing exactly one isolated charged lepton (electron or muon)
with $\pT > \SI{30}{\GeV}$ and $|\eta|<2.5$ are accepted.
Candidate events must have exactly two jets satisfying the criteria described
in Section~\ref{sec:obj}. Jets reconstructed in the range $2.75<|\eta|<3.5$,
covering the endcap--forward calorimeter transition region, must have $\pT >\SI{35}{\GeV}$.
At least one of the selected jets is required to be identified ($b$-tagged) 
as a $b$-jet. 

In order to reduce the number of multijet background events, which are characterised by low $\MET$ and 
low $W$-boson transverse mass\footnote{The $W$-boson transverse mass is defined as:
$\mtw~=~\sqrt{ 2 \left[\pT(\ell) \MET  - \vec{p}_{\mathrm T}(\ell) \cdot
\vec{E}_{\mathrm{T}}^{\mathrm{miss}}\right]}$, where $\vec{p}_{\mathrm T}(\ell)$ denotes the transverse momentum of the
electron or muon and $\pT(\ell)$ its modulus.} \mtw, the event selection requires $\MET > \SI{30}{\GeV}$ and $\mtw >
\SI{50}{\GeV}$.
To further suppress the multijet background
a requirement on the \pT of the charged lepton and the azimuthal angle
between the charged lepton and jet is applied: 
\begin{equation}
\pT\left(\ell\right) > \max\left(\SI{30}{\GeV}, \SI{40}{\GeV} \cdot \frac{|\Delta \phi\left(j_1, \ell
\right)|}{\pi}\right),
\end{equation}
where $\ell$ denotes the identified charged lepton and $j_1$ the reconstructed jet
with the highest $\pT$.

Contributions from processes with two isolated leptons in the
final state are suppressed by rejecting any event with an additional electron or muon as defined
above satisfying \pT > \SI{10}{\GeV}.

Three kinematic regions are defined in this analysis, all three being subject to the same event selection
requiring one electron or muon, missing transverse momentum and one or two $b$-tagged jets: 
\begin{itemize}
  \item The signal region (SR) is defined by using the default $b$-tagging requirement 
        and selecting exactly one $b$-tagged jet.
  \item The $W$-boson validation region (\wcr) requires exactly one $b$-tagged jet, but with a less stringent
  $b$-tagging requirement with a $b$-tagging efficiency of \SI{85}{\%}. Events contained in the SR are rejected.
  The validation region is defined such, that the composition
  of the resulting sample is dominated by $W$+jets production with a purity of \SI{77}{\%} and the same
  reconstruction of the top-quark kinematics can be used as in the signal region, in order to check the modelling of
  kinematic variables.
 \item Events in the $\ttbar$ validation region (\tcr) are required to have exactly three jets of which exactly two are
 $b$-tagged jets using the default $b$-tagging requirement. This validation region is highly enriched in $\ttbar$
       events with a purity of \SI{85}{\%}.
\end{itemize}

\section{Background estimation}
\label{sec:background}
For all background processes, except the multijet background, 
the number of expected events are obtained from Monte Carlo (MC)
simulation scaled to the theoretical cross-section predictions.
%
The associated production of an on-shell $W$ boson and a top quark ($Wt$) has a 
predicted production cross-section of \SI{71.1}{pb}~\cite{Kidonakis:2010ux} calculated
at approximate NNLO accuracy. 
Predictions of the $s$-channel production are calculated at NLO using the same
methodology as for the $t$-channel production and yield a cross-section of \SI{10.3}{pb}. 
The predicted \ttbar cross-section is $\sigma_{\ttbar} = \SI{831.8}{pb}$. It has been calculated at
NNLO in QCD including resummation of next-to-next-to-leading logarithmic (NNLL) soft gluon terms with 
top++2.0~\cite{Cacciari:2011hy,Baernreuther:2012ws,Czakon:2012zr,Czakon:2012pz,Czakon:2013goa,Czakon:2011xx}{}.
All quoted cross-sections are given for $\mtop = \SI{172.5}{\GeV}$.
The inclusive cross-sections of $W$+jets and $Z$+jets production are calculated at NNLO 
with \textsc{FEWZ}~\cite{Anastasiou:2003ds}. 
Diboson events are normalised to the NLO cross-section provided by the \SHERPA generator.

The matrix method~\cite{ATLAS-CONF-2014-058} is used to determine the multijet background 
in the electron channel.
This method estimates the number of multijet background events in the signal region
by applying efficiency factors to the number of events passing the signal tight and a loose lepton selection,
the former selection being a subset of the latter.
The number of multijet events $N^\mathrm{tight}_\mathrm{fake}$ passing the signal requirements can be expressed as
\begin{equation}
N^\mathrm{tight}_\mathrm{fake} = \frac{\epsilon_\mathrm{fake}}{\epsilon_\mathrm{real} - \epsilon_\mathrm{fake}} \cdot (N^\mathrm{loose} \epsilon_\mathrm{real} - N^\mathrm{tight}),
\label{eqn:intro-mm-tight_fake}
\end{equation}
where $\epsilon_\mathrm{real}$ and $\epsilon_\mathrm{fake}$ are the efficiencies for real and fake loose leptons being
selected as tight leptons, $N^\mathrm{loose}$ is the number of selected events in the loose sample, and
$N^\mathrm{tight}$ is the number of selected events in the signal sample.
The fake-lepton efficiencies are determined from a data sample dominated
by non-prompt and fake-lepton background events. This sample is selected by
requiring exactly one loose lepton and low \MET as well as low \mtw. The
real-lepton efficiencies are also estimated from collision data using a ``tag-and-probe'' method in $Z \rightarrow ee$ events.

Multijet-background events containing non-prompt muons are modelled with a sample 
of events enriched in non-isolated muons~\cite{ATLAS-CONF-2014-058}. Most of
these events originate from $b$-hadron or $c$-hadron decays in jets. These events pass the same kinematic 
requirements as the events of the signal sample. Only some of the muon identification 
cuts are modified, ensuring that there is no overlap with the signal selection.
The normalisation is determined using a binned maximum-likelihood fit.
 
The fit is performed to the observed data in the \mtw distribution after applying all selection criteria,
except the requirement on \mtw.
The multijet template is fit together with templates derived from MC simulation for all other 
processes. The rate uncertainties are accounted for in the fitting process 
in the form of additional constrained nuisance parameters. For the purpose of this fit,
three different template distributions are used. 
One template is built from simulated $W$+jets events, one consists of events from $\ttbar$ and single
top-quark production, and one consists of contributions from $Z$+jets and $VV$ production.
As the shape of the joint template of $Z$+jets and $VV$ events is very similar to that of $W$+jets events, 
the rates are fixed in the fitting process. 

The estimated event rates obtained from the binned maximum-likelihood fit for the combined contributions 
of $W$+jets, \ttbar and single top-quark production are not used in the later analysis
and are only applied to scale the respective processes in order to check the modelling of the
kinematic distributions. For the neural-network training, as well as for the final statistical analysis,
the normalisation for all but the multijet background is taken from MC simulations scaled 
to their respective cross-section predictions.

In the signal region, 34459 events in the \lpc and 31056 events in the \lmc are observed 
in data, while the expected SM background amounts to $33600 \pm 2600$ events and
$30200 \pm 2300$ events, respectively. The quoted uncertainties are statistical
uncertainties and the uncertainty in the number of multijet
events. Table~\ref{tab:yields} summarises the event yields in the signal region for each of the
background processes considered together with the event yields for the signal process. 
The yields are calculated using the acceptance from MC samples
normalised to their respective theoretical cross-sections including the (N)NLO $K$ factors.

\begin{table}[htb]
\sisetup{round-mode=figures}
\begin{center}
\begin{tabular}{
                 l S[table-format=5.0,round-precision=3]@{$\,\pm\,$}S[table-format=4.0,round-precision=2]
                   S[table-format=5.0,round-precision=3]@{$\,\pm\,$}S[table-format=4.0,round-precision=2]
}
\hline
    \toprule
    Process & \multicolumn{2}{c}{\lpc} &  \multicolumn{2}{c}{\lmc} \\ 
    \midrule
    $tq$        & 4195 &  168 &  8  &  3    \\
    $\bar{t}q$  &  5  &  2  & 2705 & 137     \\
    \midrule
    $t\bar{t}$  & 13050 & 791 & 13078 & 793      \\
    $Wt$        &  1641 & 106 &  1643 & 107      \\
    $t\bar{b}$+$\bar{t}b$ & 298 & 25 & 199 & 18 \\
    $W^+$+jets    & 10472 & 2160  & \multicolumn{2}{c}{<1}     \\
    $W^-$+jets    & \multicolumn{2}{c}{<1} & 8726 & 1800     \\
    $Z,VV$+jets  & 1531 & 324 & 1406 & 298  \\
    Multijets   & 2400 & 1211 & 2400 & 1211 \\ 
    \midrule
    Total expected & 33613 & 2627 & 30184 & 2336  \\
    Data observed & \multicolumn{2}{l}{\num[round-mode=off]{34459}} & \multicolumn{2}{l}{\num[round-mode=off]{31056}}  \\
    \bottomrule
\end{tabular}
\caption{\label{tab:yields} Predicted and observed event yields for the signal region.
The quoted uncertainties include uncertainties in the theoretical cross-sections, in the number of multijet events,
and the statistical uncertainties.}
\end{center}
\end{table}
In the following, the electron and muon channel are combined for all
figures and fits. Different processes are also grouped together in the following
way. The top-quark background consists of all background processes that
include the production of top quarks. These processes are \ttbar~production and single top-quark
production in the $Wt$ and $t\bar{b}$+$\bar{t}b$ channel. The $W$+jets
process describes the production of a real $W$ boson in association
with jets, while the production of a $Z$ boson or two vector bosons $VV$ in
association with jets are grouped together to $Z,VV$+jets.
Finally, multijets represents events with a fake lepton originating from
multijet production.

\clearpage

\section{Discrimination of signal and background events}
\label{sec:nn}
\begin{figure}[!ht]
\begin{center}
\subfigure[]{
 \includegraphics[width=0.455\textwidth]{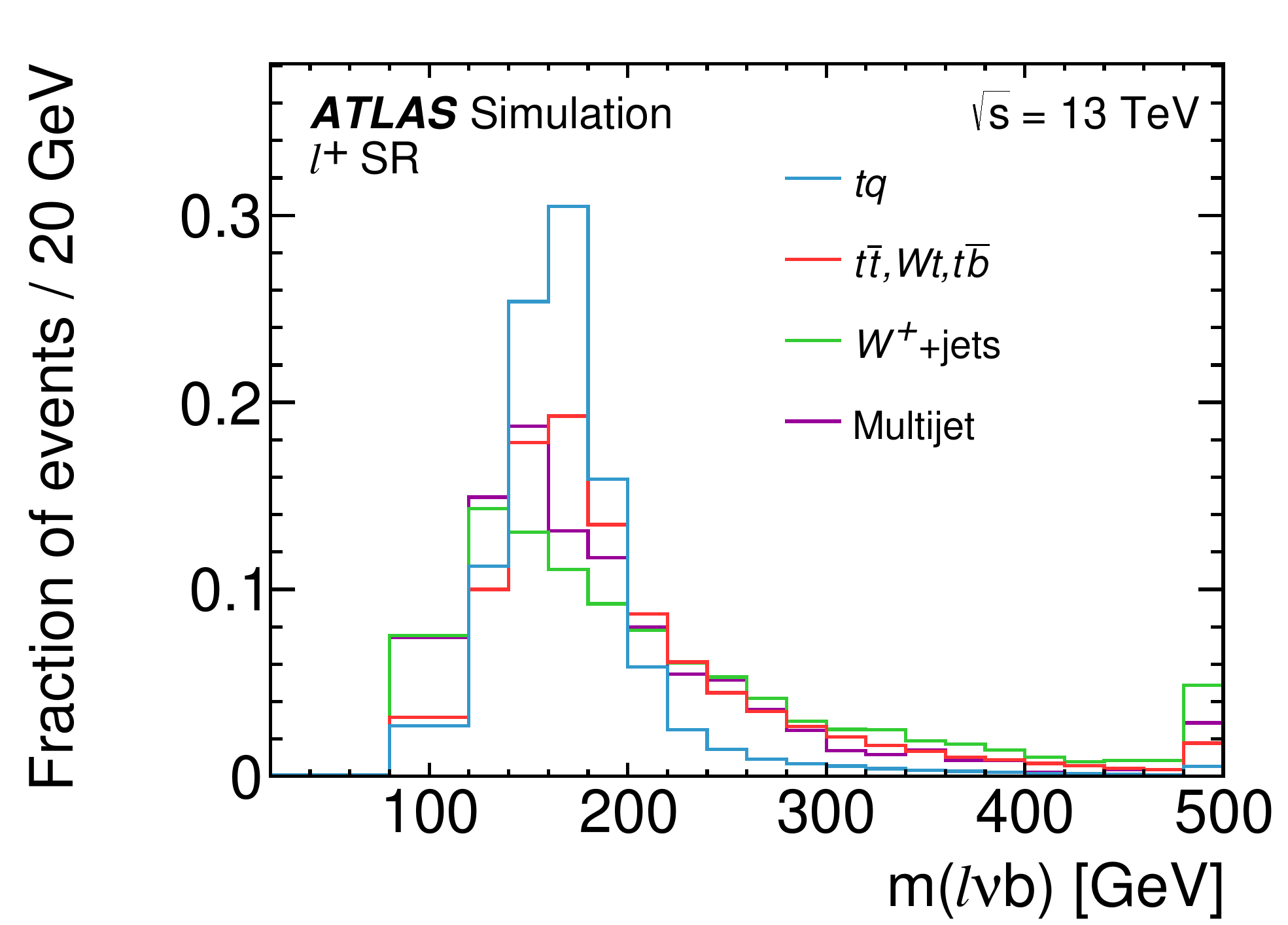}
 \label{subfig:mlnub_plus_shape}
}
\subfigure[]{
 \includegraphics[width=0.455\textwidth]{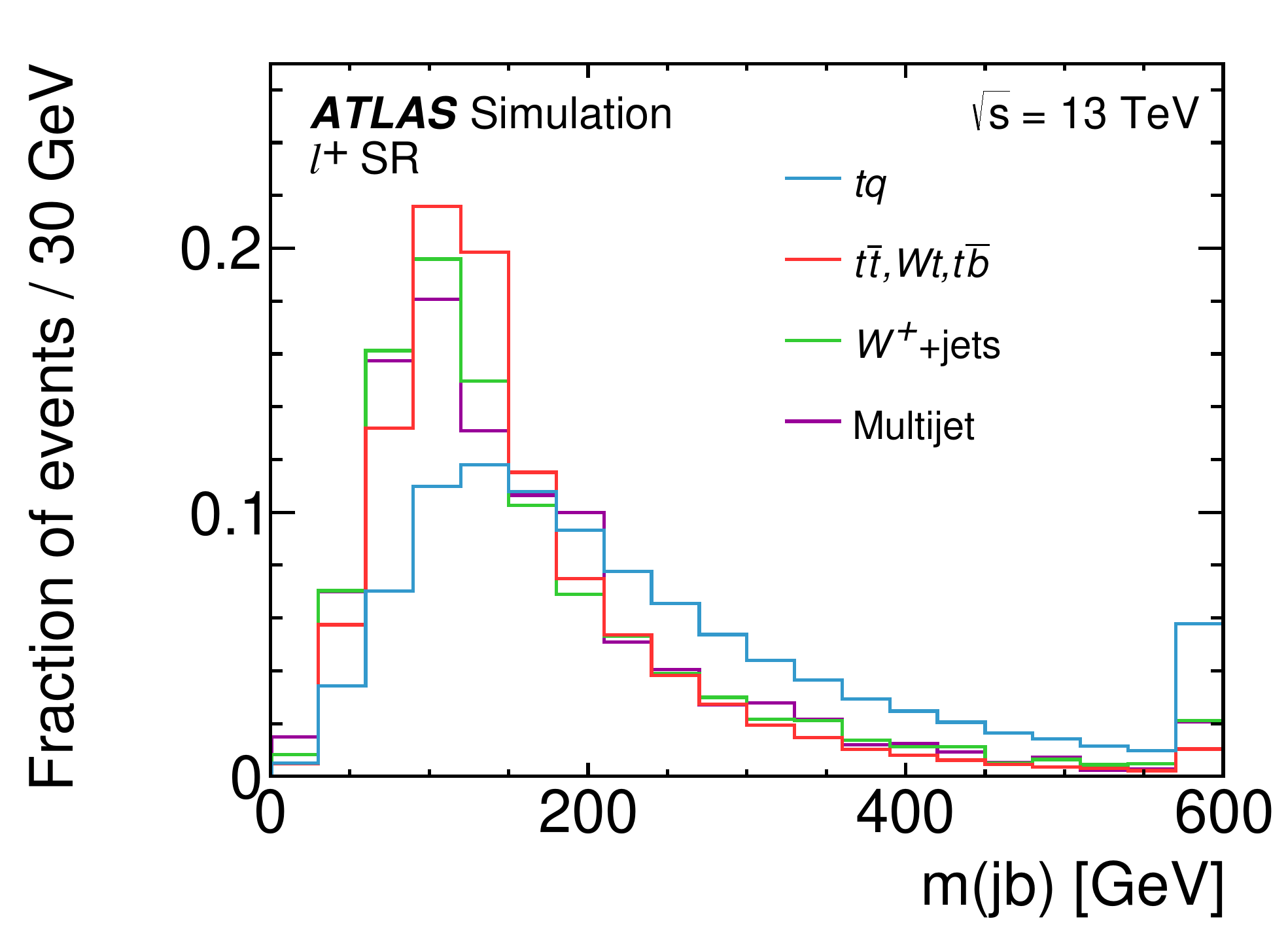}
 \label{subfig:mj1j2_plus_shape}
}
\subfigure[]{
\includegraphics[width=0.455\textwidth]{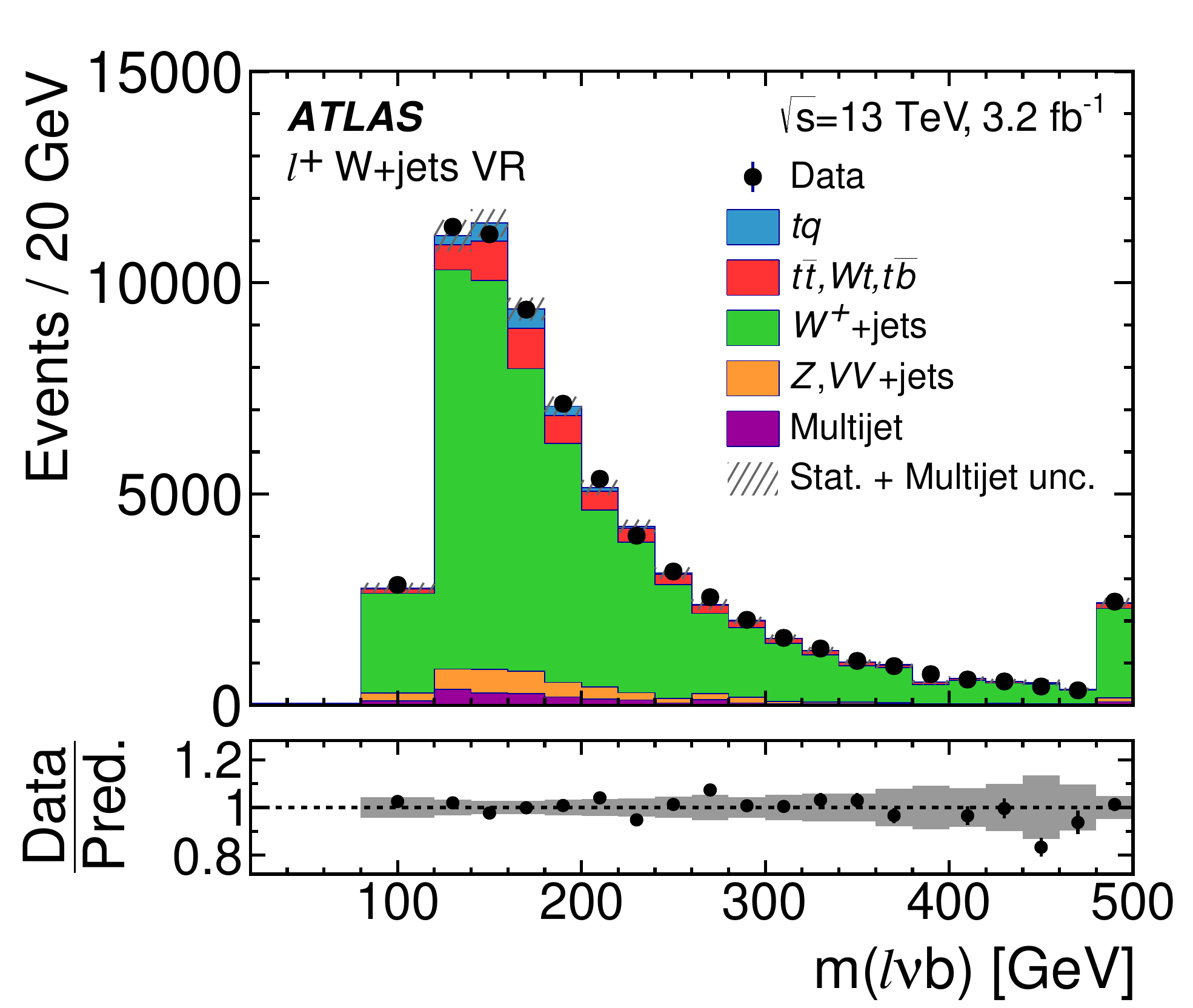}
\label{subfig:mlnub_plus_wcr}
}
\subfigure[]{
\includegraphics[width=0.455\textwidth]{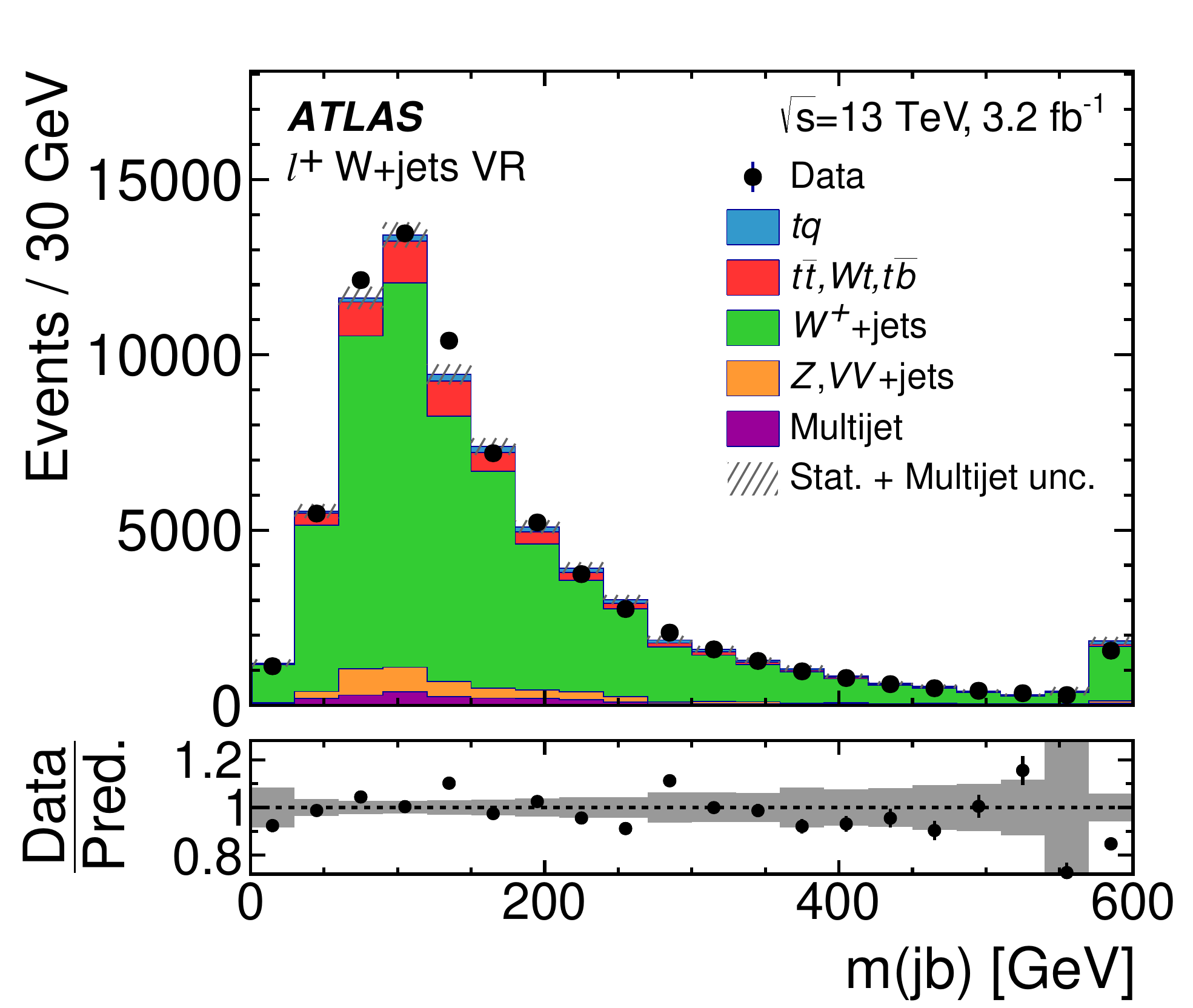}
\label{subfig:mj1j2_plus_wcr}
}
\subfigure[]{
\includegraphics[width=0.455\textwidth]{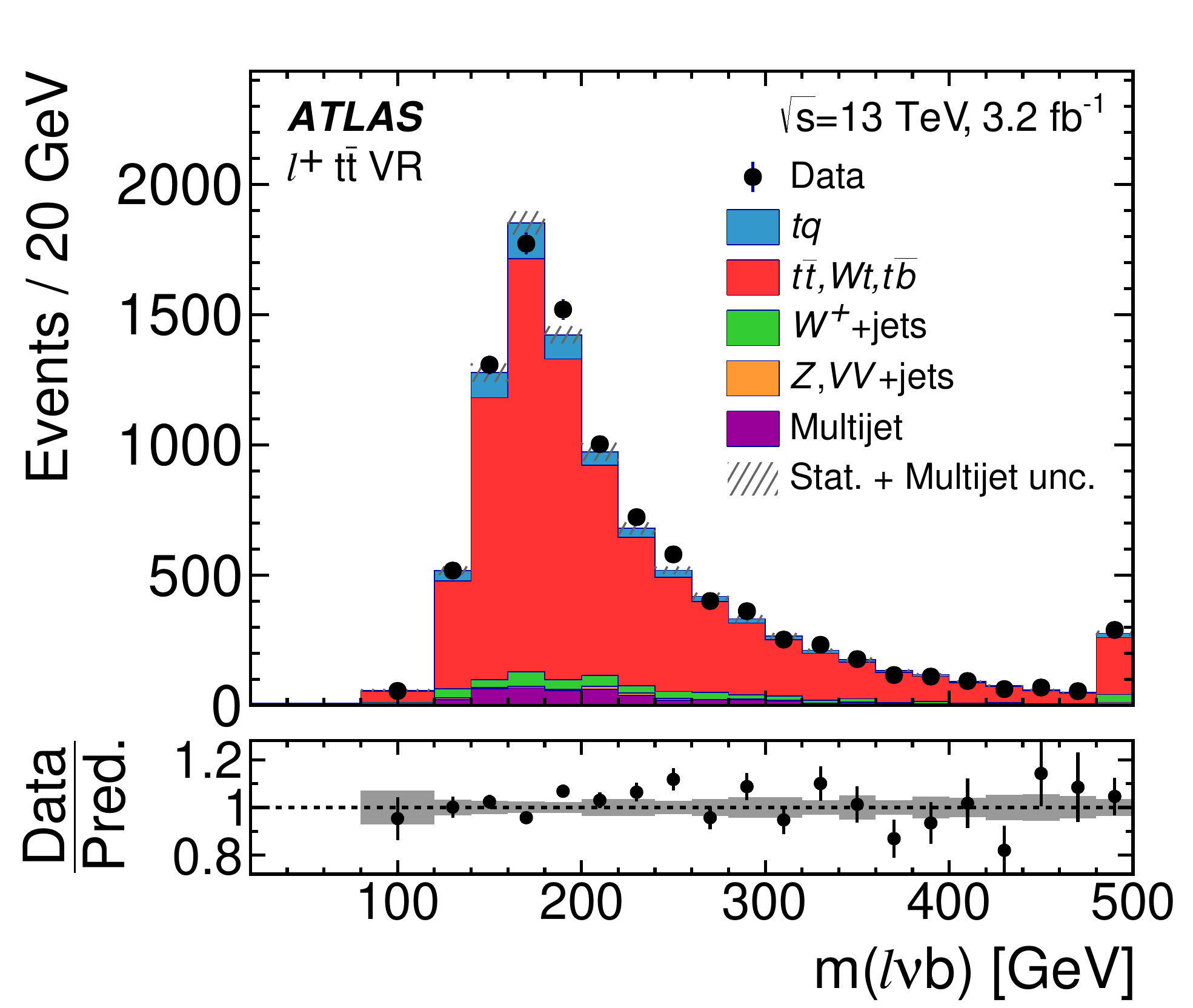}
\label{subfig:mlnub_plus_tcr}
}
\subfigure[]{
\includegraphics[width=0.455\textwidth]{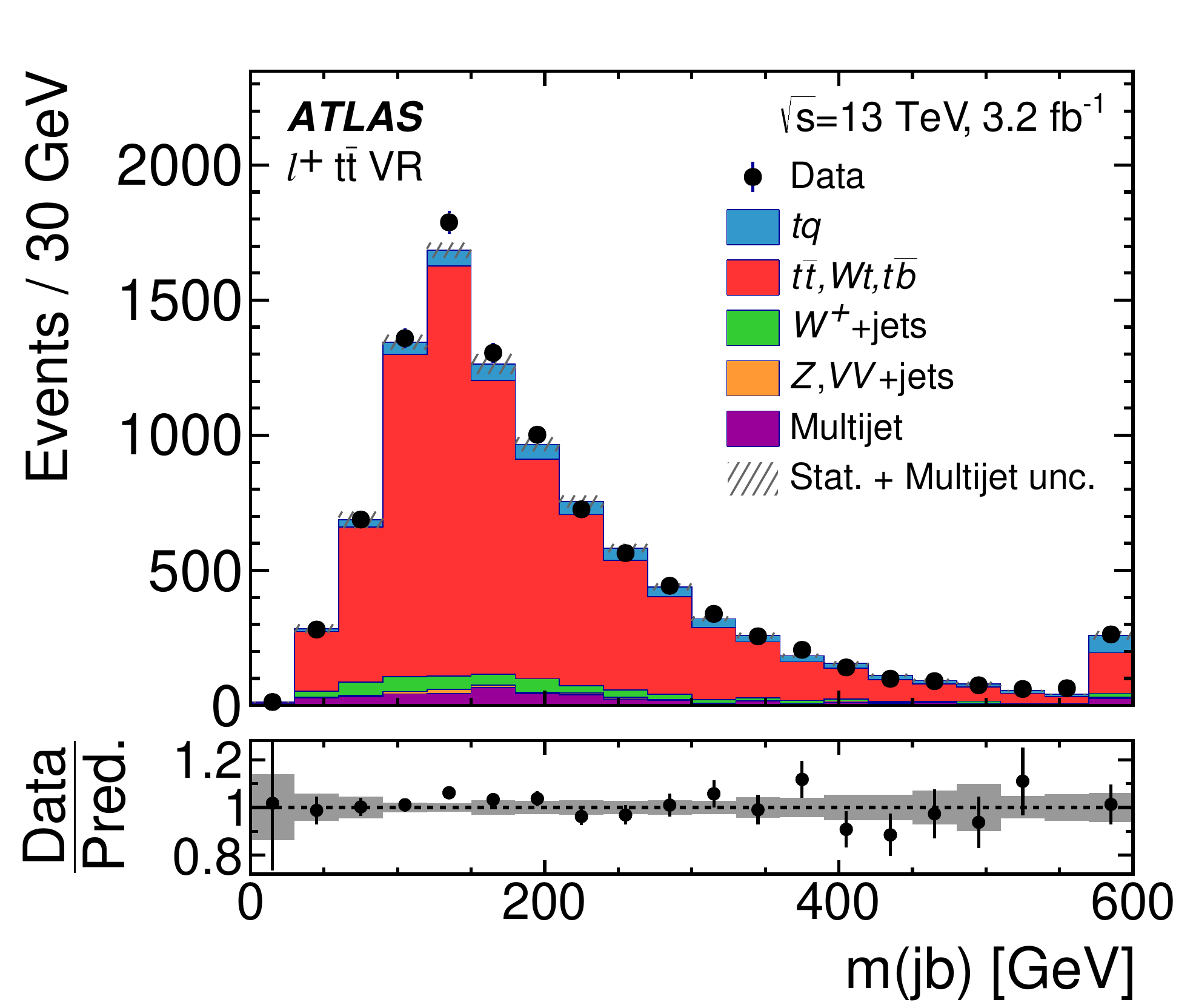}
\label{subfig:mj1j2_plus_tcr}
}
\vspace{-0.5cm}
 \caption{
 Distributions of the two most discriminating variables, (left) the reconstructed top-quark mass $m(\ell\nu b)$
 and (right) the invariant mass of the jet pair $m(j b)$, for the \lpc.
 In the \tcr the $b$-jet used to calculate $m(l\nu b)$ and $m(jb)$ is the
 $b$-jet with the higher \pT.
\subref{subfig:mlnub_plus_shape}-\subref{subfig:mj1j2_plus_shape}:
Signal and background distributions normalised to unit area.
\subref{subfig:mlnub_plus_wcr}-\subref{subfig:mj1j2_plus_tcr}: Observed 
distributions in the \wcr and the \tcr
compared to the model obtained from simulated events.
The simulated distributions are 
normalised to match the number of observed events as described in the main text.
  The hatched and grey error bands represent the uncertainty in the number of multijet events and the
  uncertainty due to the size of the MC samples. 
  The ratio of observed to predicted (Pred.) number of events in
  each bin is shown in the lower distributions. 
  Events in the overflow are contained in the last bin.
}
 \label{fig:twomost}
\end{center}
\end{figure}

To separate $t$-channel single-top-quark signal events from the background, several kinematic variables are combined
into one discriminant by employing a neural network~\cite{feindt-2004,Feindt:2006pm}. 
A large number of potential input variables were 
studied, including kinematic variables of the identified physics objects, as well as
variables obtained from the reconstruction of the $W$ boson and the top quark.
A detailed description of the algorithm including the reconstruction of 
the longitudinal component of the neutrino momentum is given in Ref.~\cite{TOPQ-2012-21}.
As a compromise between the discrimination power and the need for a manageable number of 
variables, the ten highest-ranking variables are chosen and are listed in Table~\ref{tab:trainingVars}.
The two most discriminating variables are the reconstructed top-quark mass $m(\ell \nu b)$ and the 
invariant mass of the two jets $m(j b)$. 
Figures~\ref{fig:twomost}\subref{subfig:mlnub_plus_shape} and \ref{fig:twomost}\subref{subfig:mj1j2_plus_shape}
show the $m(l\nu b)$ and $m(jb)$ distributions (normalised to unit area) in the SR for the $\ell^{+}$ channel.
Figures~\ref{fig:twomost}\subref{subfig:mlnub_plus_wcr}--\ref{fig:twomost}\subref{subfig:mj1j2_plus_tcr} show 
the $m(l\nu b)$ and $m(jb)$ distributions in the \wcr and \tcr for the $\ell^{+}$ channel.
In the \tcr the $b$-jet used to calculate $m(l\nu b)$ and $m(jb)$ is the
$b$-jet with the higher \pT. The distributions from the different
processes, apart from the multijet background in the electron channel, are normalised to match the number of observed events.
In the case of the electron channel, the relative contribution of each simulated
process is estimated using its predicted cross-section. In the case of the muon
channel, the distributions are normalised to the expected number of events obtained from the fit to the $\mtw$ distributions described in
Section~\ref{sec:background}. Satisfactory agreement is seen between the data and the predictions.

\begin{figure}
\begin{center}
\subfigure[]{
 \includegraphics[width=0.455\textwidth]{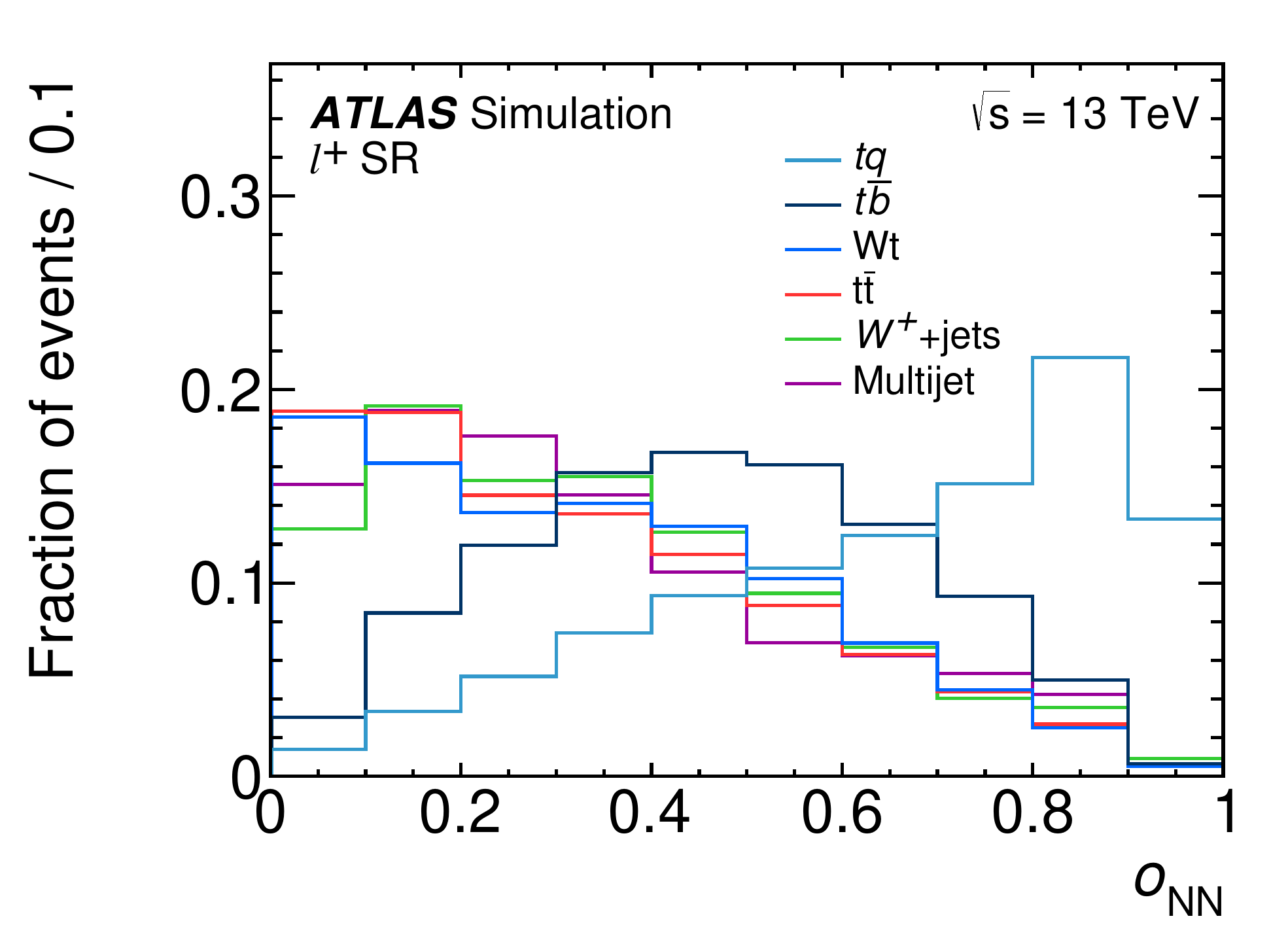}
 \label{subfig:nn_plus_shape}
} 
\subfigure[]{
 \includegraphics[width=0.455\textwidth]{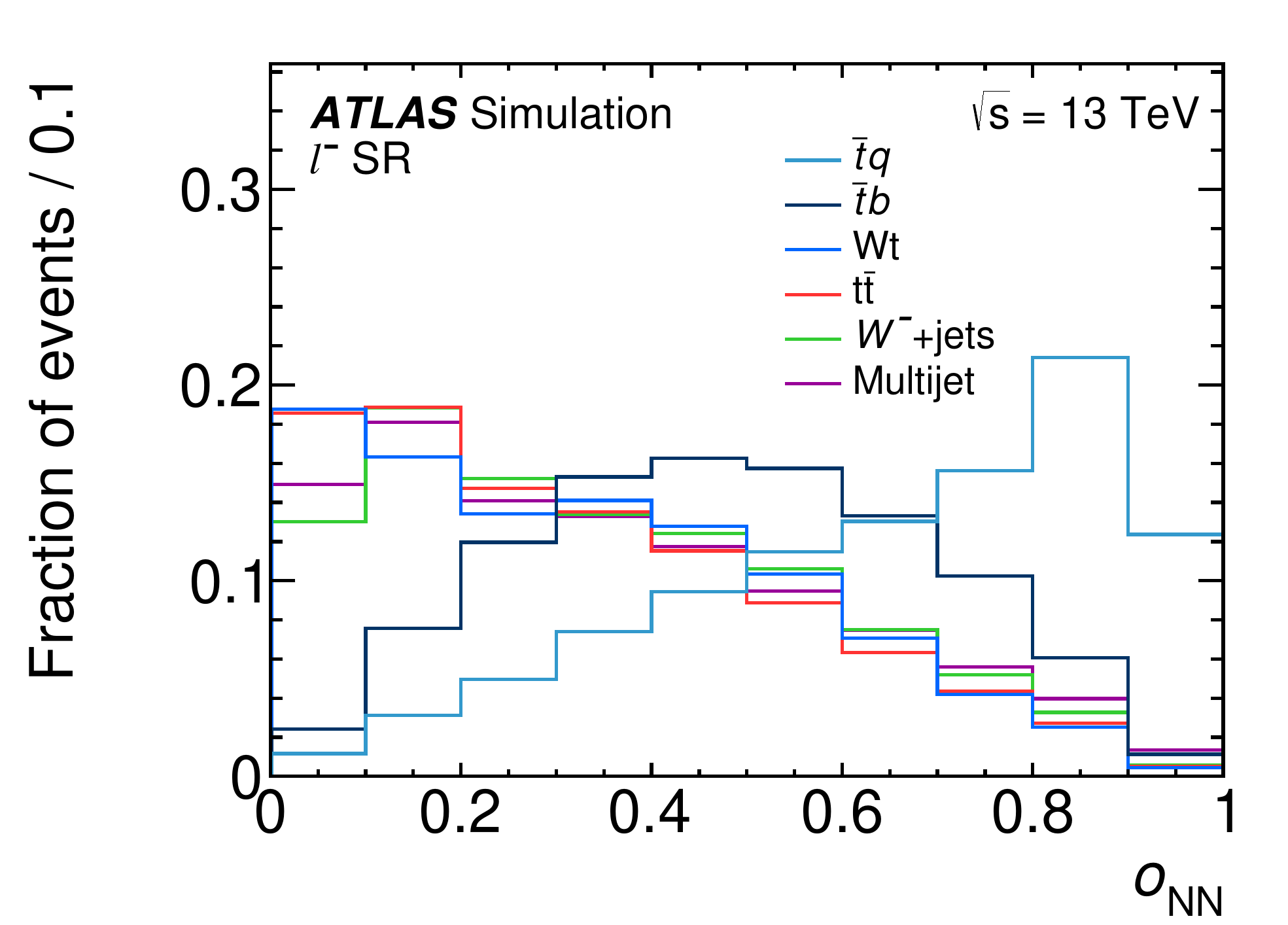}
 \label{subfig:nn_minus_shape}
} 
\subfigure[]{
\includegraphics[width=0.455\textwidth]{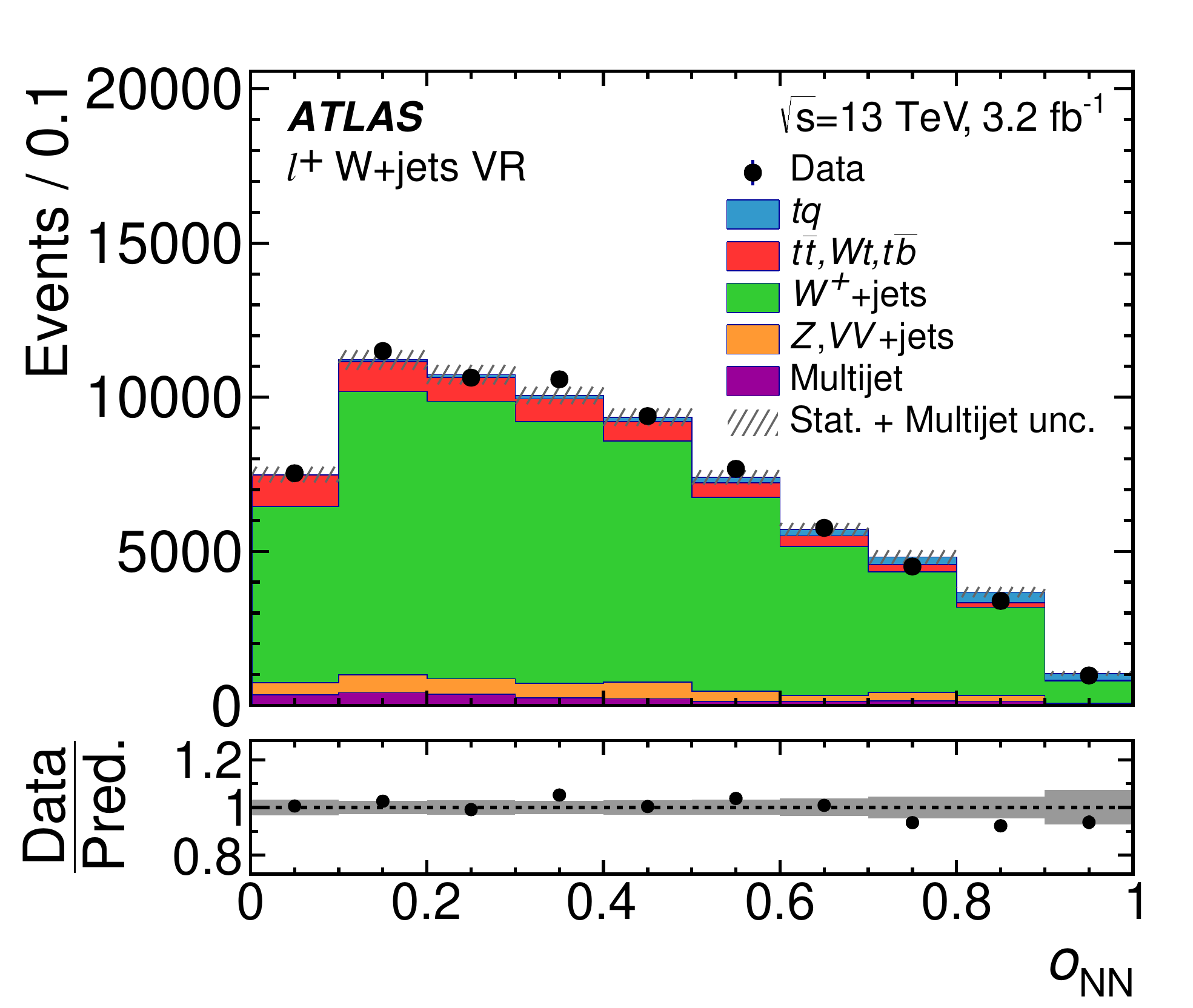}
\label{subfig:nn_plus_wcr}
}
\subfigure[]{
\includegraphics[width=0.455\textwidth]{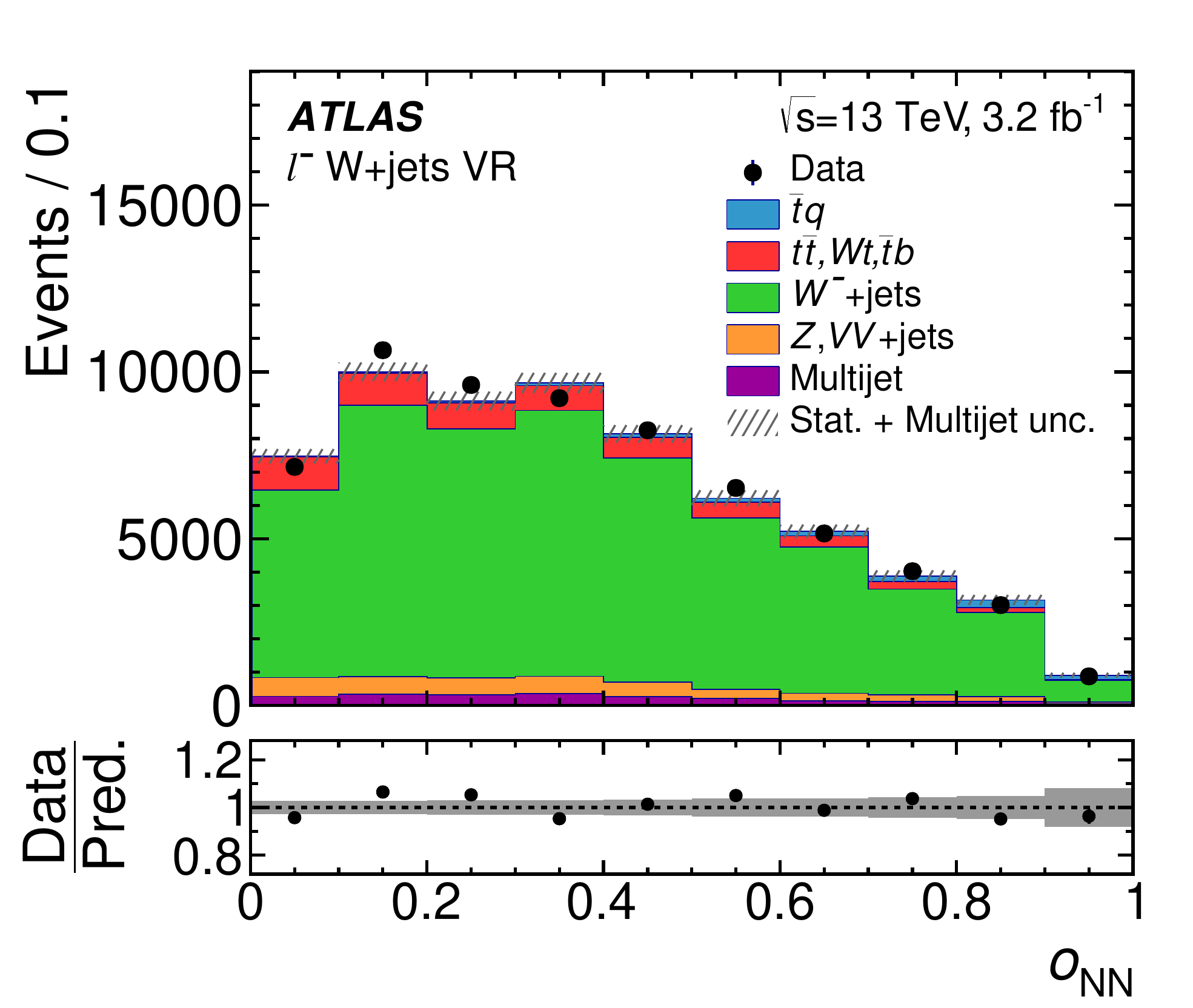}
\label{subfig:nn_minus_wcr}
}
\subfigure[]{
\includegraphics[width=0.455\textwidth]{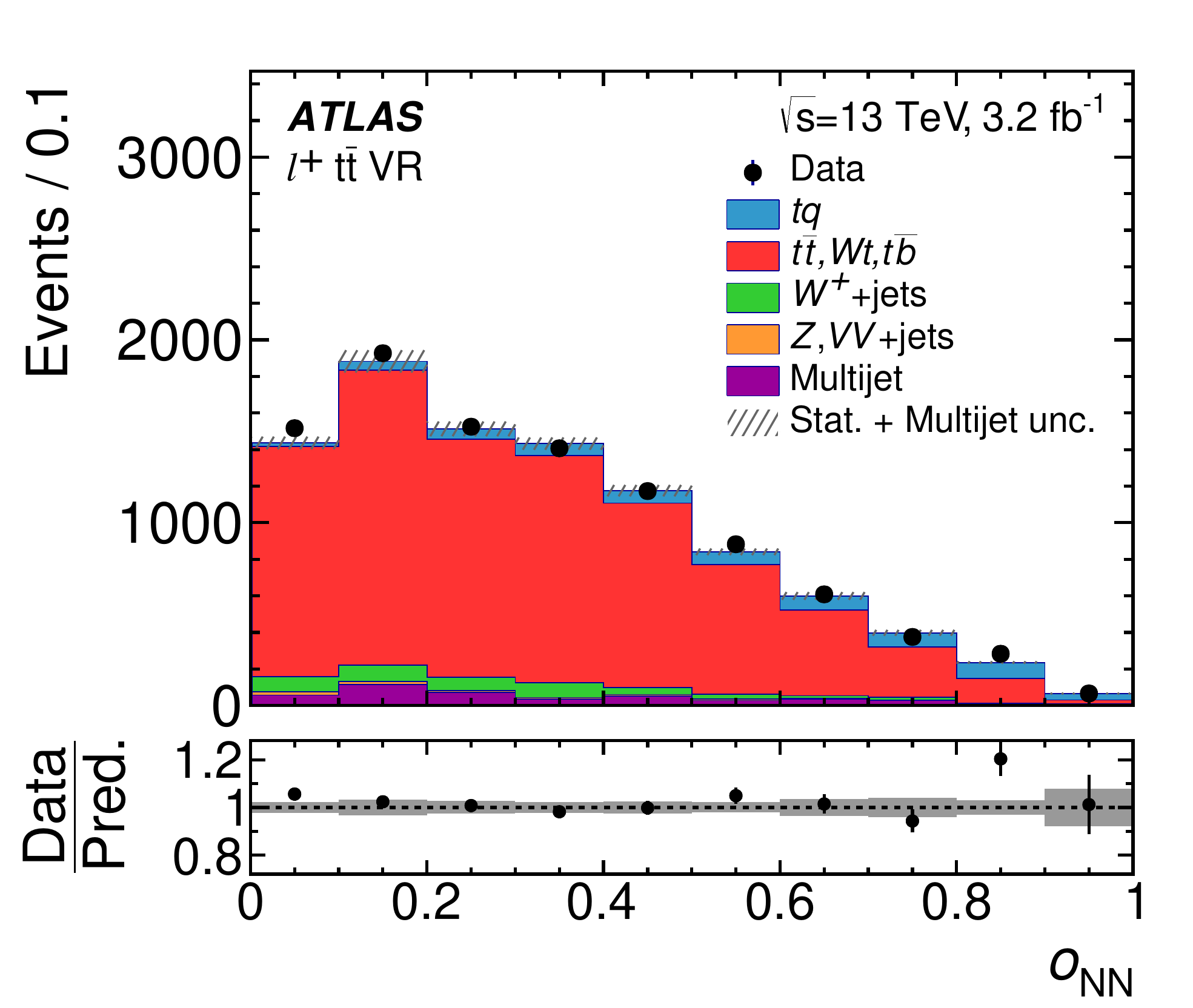}
\label{subfig:nn_plus_tcr}
}
\subfigure[]{
\includegraphics[width=0.455\textwidth]{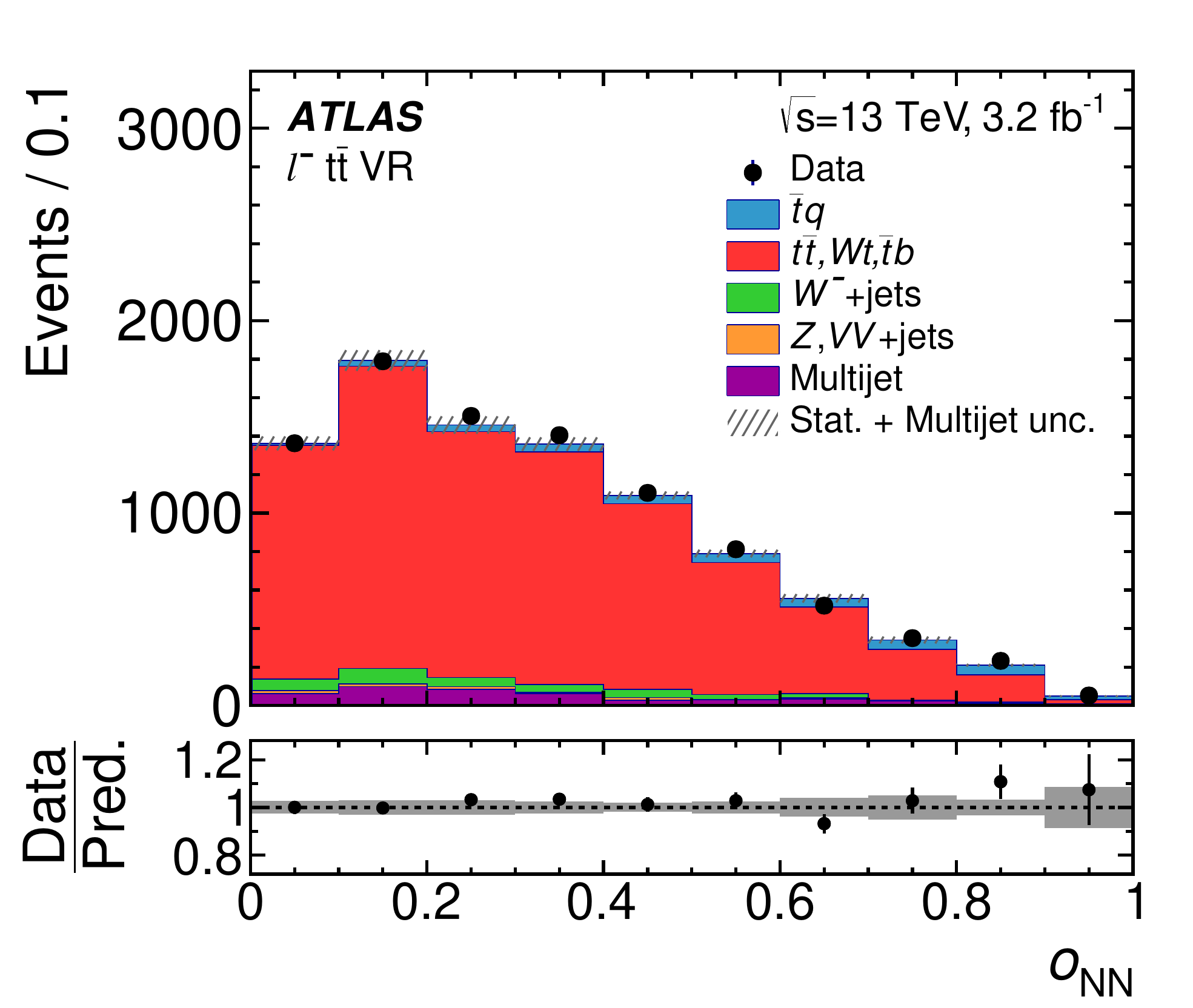}
\label{subfig:nn_minus_tcr}
}
\vspace{-0.5cm}
\caption{\label{fig:nn2jets}
Distributions of the NN discriminant $O_{\mathrm{NN}}$ (left) for the \lpc and (right) for the \lmc.
\subref{subfig:nn_plus_shape}-\subref{subfig:nn_minus_shape}: Signal and background distributions normalised to unit area.
\subref{subfig:nn_plus_wcr}-\subref{subfig:nn_minus_tcr}:  Observed 
distributions in the \wcr and the \tcr
compared to the model obtained from simulated events.
The simulated distributions are 
normalised to match the number of observed events as described in the main text.
  The hatched and grey error bands represent the uncertainty in the number of multijet events and the uncertainty 
  due to the size of the MC samples. 
  The ratio of observed to predicted (Pred.) number of events in
  each bin is shown in the lower distributions.
}
\end{center}
\end{figure}

The NN infrastructure consists of one input node for each input
variable plus one bias node, an arbitrary user-defined number of hidden nodes,
and one output node which gives a continuous output in the interval [$0,1$]. In
this specific case, 15 nodes in the  hidden layer are used 
and equal numbers of signal and background events were used in the training, where the
different background processes are weighted according to their expected number of events.
The shapes of the resulting NN discriminant distributions ($O_{\mathrm{NN}}$) for the signal and the two largest 
backgrounds are shown in Figure~\ref{fig:nn2jets} together with the data distributions compared to the
predictions in the two validation regions. Good agreement between the predicted and observed 
distributions is found.
\begin{table}[ht!]
\begin{center}
\begin{tabular}{ll}
\hline
Variable & Definition\\
\hline
    $m(\ell \nu b)$  & top-quark mass reconstructed from the charged lepton,\\
                     & neutrino, and $b$-tagged jet \\
    $m(j b)$   & invariant mass of the $b$-tagged and untagged jet\\
    \mtw       & transverse mass of the reconstructed $W$ boson\\
    $|\eta(j)|$  & modulus of the pseudorapidity of the untagged jet\\
    $m(\ell b)$  &  invariant mass of the charged lepton ($\ell$) and the $b$-tagged jet\\
    $\eta(\ell \nu)$  & rapidity of the reconstructed $W$ boson\\
    $\Delta R(\ell \nu b,j)$ & $\Delta R$ of the reconstructed top quark and the untagged jet\\
    $\cos \theta^*(\ell,j)$  & cosine of the angle $\theta^*$ between the charged
    lepton and the untagged jet\\
                     & in the rest frame of the reconstructed top quark \\
    $\Delta \pT(\ell \nu b,j)$ & $\Delta \pT$ of the reconstructed top quark and the untagged jet\\
        
    $\Delta R(\ell,j)$ & $\Delta R$ of the charged lepton and the untagged jet\\
\hline
\end{tabular}
  \caption{\label{tab:trainingVars} The ten variables that are used in the training of the neural network ordered by
  their discriminating power as determined by
  Neurobayes~\cite{Feindt:2006pm,TOPQ-2012-21}.}
\end{center}
\end{table}
\clearpage

\section{Systematic uncertainties}
\label{sec:systematics}
Systematic uncertainties in the normalisation of the individual
backgrounds and in the signal acceptance as well as uncertainties in the 
shape of the NN discriminant distribution of the individual predictions 
affect the individual top-quark and top-antiquark 
cross-section measurements and their ratio. 
The uncertainties are split into the following categories.

\vspace{-0.3cm}
\paragraph{Reconstruction efficiency and calibration uncertainties}
Systematic uncertainties
affecting the reconstruction and energy calibration of 
jets, electrons, and muons are propagated through the analysis.
The dominant source for this measurement arises from the jet energy scale (JES) calibration, 
including the modelling of pile-up, and from the $b$-jet tagging efficiencies. 

The uncertainties due to lepton reconstruction, 
identification and trigger efficiencies are estimated using tag-and-probe methods in $Z \rightarrow \ell\ell$
events.
Correction factors are derived to match the simulation to observed
distributions in collision data and associated uncertainties are estimated.
To estimate uncertainties in the lepton momentum scale and resolution,
also $Z \rightarrow \ell\ell$ events are used~\cite{PERF-2013-05,ATL-PHYS-PUB-2015-041,PERF-2015-10}.
The lepton-charge misidentification is estimated with simulated events and found
to be below \SI{0.1}{\%}, see Table~\ref{tab:yields}.
The uncertainty on the lepton-charge misidentification is evaluated and found to be negligible.

Several components of the JES uncertainty are considered~\cite{PERF-2012-01,ATL-PHYS-PUB-2015-015}.
Uncertainties derived from different dijet-\pT-balance measurements as well as 
uncertainties associated with in-situ calibration techniques are considered.
Furthermore, the presence of nearby jets and the modelling of pile-up affects the jet calibration.
The uncertainty in the flavour composition covers effects due to the difference in quark--gluon
composition between the jets used in the calibration and the jets used in this analysis. 
Also an uncertainty due to limited knowledge of the calorimeter response to light-quark
jets and gluon jets is considered.
Finally, the JES uncertainty is estimated for $b$-quark jets by varying the modelling of $b$-quark
fragmentation.
The uncertainty in the jet energy resolution is modelled
by varying the \pT~of the jets according to the systematic uncertainties of the resolution 
measurement~\cite{PERF-2011-04}.
The effect of uncertainties associated with the JVT requirement is also considered.

The impact of a possible miscalibration on the soft track component
of \MET{} is derived from data--MC comparisons of the \pT balance
between the hard and soft \MET components.

Since the analysis makes use of $b$-tagging, the uncertainties in the $b$- and $c$-tagging 
efficiencies and the mistag rate are taken into account.  
These uncertainties were determined using $\sqrt{s}= \SI{8}{\TeV}$ data as described
in Ref.~\cite{ATLAS-CONF-2014-004} for $b$-jets and Ref.~\cite{ATLAS-CONF-2014-046}
for $c$-jets and light jets, with additional uncertainties to account for the presence
of the newly added insertable B-layer
and the extrapolation to $\sqrt{s}= \SI{13}{\TeV}$.

\vspace{-0.3cm}
\paragraph{Monte Carlo generators}
Systematic effects from MC modelling of the signal and the \ttbar background process  
are either estimated by comparing different generators or by 
comparing parameter variations in the \powheg + \PYTHIA6 setup.
The \powheg + \herwigpp sample is used for parton shower and hadronisation modelling studies,
while \MGMCatNLO + \herwigpp is used for studies of the NLO-matching method.
Variations of the amount of additional radiation are studied by changing $\mu_{\mathrm r}$ and $\mu_{\mathrm f}$
and the scales in the parton shower simultaneously. In these samples, an up-variation of 
$\mu_{\mathrm r}$ and $\mu_{\mathrm f}$ by a factor of two is combined with the 
P2012 tune with lower radiation (P2012radLo tune) than the
nominal P2012 set, and a variation of both
scales by a factor of one half is combined with the 
P2012 tune with higher radiation (P2012radHi tune).
In the case of the up-variation of \ttbar production,  the $\hdamp$ parameter is also changed
and set to two times the top-quark mass~\cite{ATL-PHYS-PUB-2015-002}.
 
The uncertainty in the pile-up reweighting as well as the 
statistical uncertainties of the simulated event samples are also taken into account.

\vspace{-0.3cm}
\paragraph{PDF}
The systematic uncertainties in the signal and background acceptance 
related to the parton distribution functions
are taken into account for all single-top-quark processes and \ttbar production.
 The procedure follows the updated PDF4LHC recommendation~\cite{Butterworth:2015oua} 
by using the 30 eigenvectors of the {\textsc PDF4LHC15} NLO PDF set.
The events are reweighted according to each of the PDF uncertainty eigenvectors.
In addition, the acceptance difference between {\textsc PDF4LHC15} and \ct is considered, since 
the latter PDF set is used in the MC samples and is not covered 
by the uncertainty obtained with {\textsc PDF4LHC15} PDF sets.

\vspace{-0.3cm}
\paragraph{Background normalisation}

The \ttbar, $Wt$ and $t\bar{b}$ backgrounds are normalised to their 
theory predictions, where a combined uncertainty of \SI{6}{\%} is derived from the weighted 
average of the individual uncertainties.
The PDF- and $\alphas$-induced uncertainties for the \ttbar process are calculated using the PDF4LHC
prescription~\cite{Botje:2011sn} with the \mstw \SI{68}{\%} CL NNLO, \ct NNLO and \nnpdf PDF sets
and added in quadrature to the uncertainty due to the scale, leading to 
a total uncertainty of \SI{5.5}{\%}. 
The uncertainty in the $Wt$ cross-section, calculated at approximate NNLO,
is the sum in quadrature of the effects of the PDF uncertainty 
obtained using the \mstw \SI{68}{\%} CL NNLO PDF sets 
and the scale uncertainty, and is found to be \SI{5.4}{\%}.
The $s$-channel production cross-section is calculated 
at NLO with a total uncertainty of \SI{3.8}{\%}.

For the $W$+jets and $Z$+jets backgrounds, an uncertainty of \SI{21}{\%} is assigned. 
This uncertainty is estimated based on parameter variations in the generation of the \SHERPA samples.
It was found that correlated variations of the factorisation and renormalisation scales have
the biggest impact on the kinematic distributions and produces change covering the 
unfolded data and their uncertainties~\cite{STDM-2012-24}.

Diboson processes have an uncertainty of \SI{6}{\%} in the inclusive cross-section including
uncertainties on the choice of the factorisation and renormalisation scales and the PDF uncertainty.

The multijet background estimate has an uncertainty of \SI{50}{\%},
based on comparisons of the rates obtained using alternative methods described in previous
analyses~\cite{TOPQ-2012-21,ATLAS-CONF-2014-007,ATLAS-CONF-2014-058}.

\vspace{-0.3cm}
\paragraph{Luminosity and beam energy}
The uncertainty in the integrated luminosity is $\pm \SI{2.1}{\%}$. It is derived,
following a methodology similar to that detailed in Refs.~\cite{DAPR-2011-01} and \cite{newlumi}, 
from a calibration of the luminosity scale using $x$--$y$ beam-separation scans performed in August 2015.
Given the level of precision of the measurement the uncertainty in the beam energy is negligible for this analysis.

All systematic uncertainties discussed above cause variations in the signal
acceptance, the background rates and the shape of the NN discriminant distribution. 
Both the rate and shape uncertainties are taken into account by generating correlated 
pseudo-experiments as detailed in the next section.

\section{Statistical analysis}
\label{sec:stat}

To extract the top-quark and top-antiquark inclusive cross-sections, 
a binned maximum-likelihood fit to the NN discriminant
distribution is performed in the \lpc and \lmc, treating 
$t$-channel top-quark and $t$-channel top-antiquark production as independent processes.
The likelihood function used is built from Poisson probability terms 
and includes Gaussian priors to constrain
the rates of the $W$+jets and top-quark background processes; more details are given in Ref.~\cite{TOPQ-2012-21}.
The fit parameters of the likelihood function are scale factors, $\beta_i$,
that multiply the expected value of the number of events, $\nu_i$, for each
fitted process $i$. 
The background normalisation constraints are \SI{21}{\%} for $W$+jets production,
and \SI{6}{\%} for top-quark backgrounds (\ttbar, $Wt$ and $t\bar{b}+\bar{t}b$), while
the contributions from $Z$+jets, $VV$, and multijet processes are fixed to their predictions. 
The fitted rates of the $W$+jets background and the top-quark backgrounds are mainly driven 
by the background-dominated region with low $O_{\mathrm{NN}}$ values.
The cross-section ratio is subsequently computed as $R_t = \sigma(tq)/\sigma(\bar{t}q)$.

\begin{figure}[!tbh]
\begin{center}
\subfigure[]{
\includegraphics[width=0.455\textwidth]{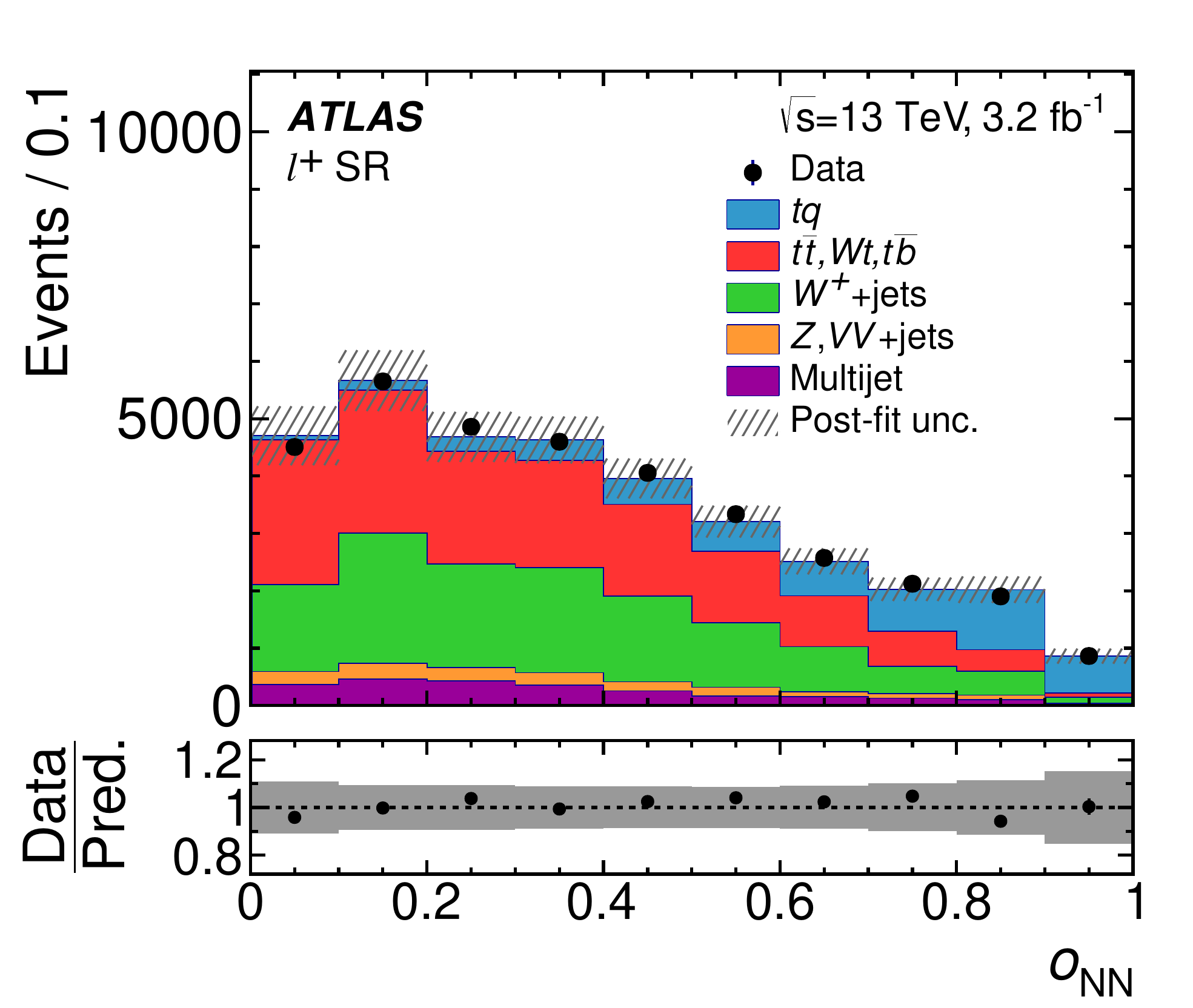}
  \label{fig:postfitnnplus}
}
\subfigure[]{
\includegraphics[width=0.455\textwidth]{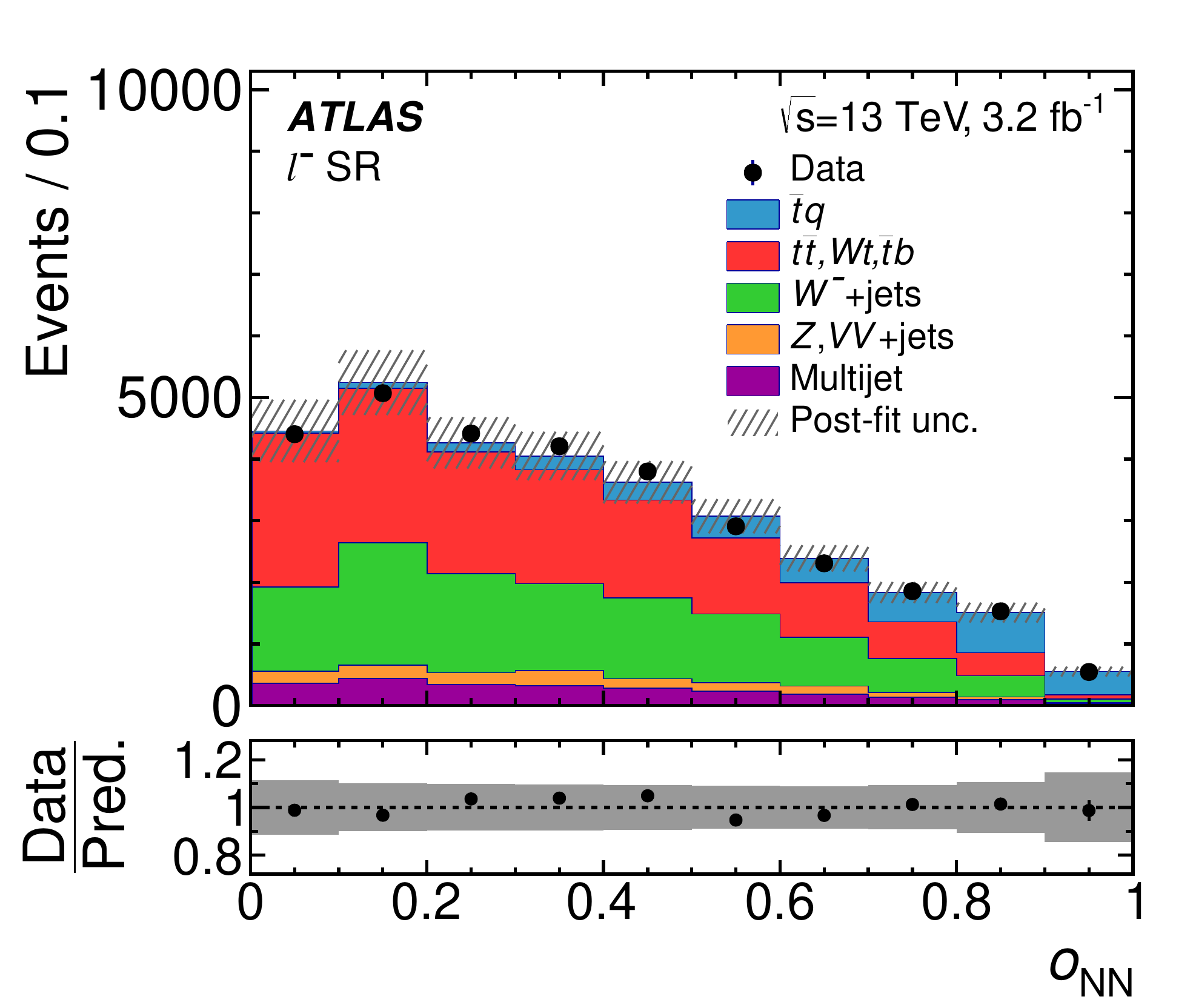}
\label{fig:postfitnnminus}
}

\caption{\label{fig:NNFit}
  NN discriminant distribution \subref{fig:postfitnnplus} for the \lpc and \subref{fig:postfitnnminus} for the \lmc in
  the SR.
  The signal and backgrounds are normalised to the fit result and 
  the hatched and grey error bands represent the post-fit uncertainty.
  The ratio of observed to predicted (Pred.) number of events in each bin is shown in the lower histogram.
 }
\end{center}
\end{figure}

\begin{figure}[!h!tpb]
\begin{center}
\subfigure[]{
\includegraphics[width=0.45\textwidth]{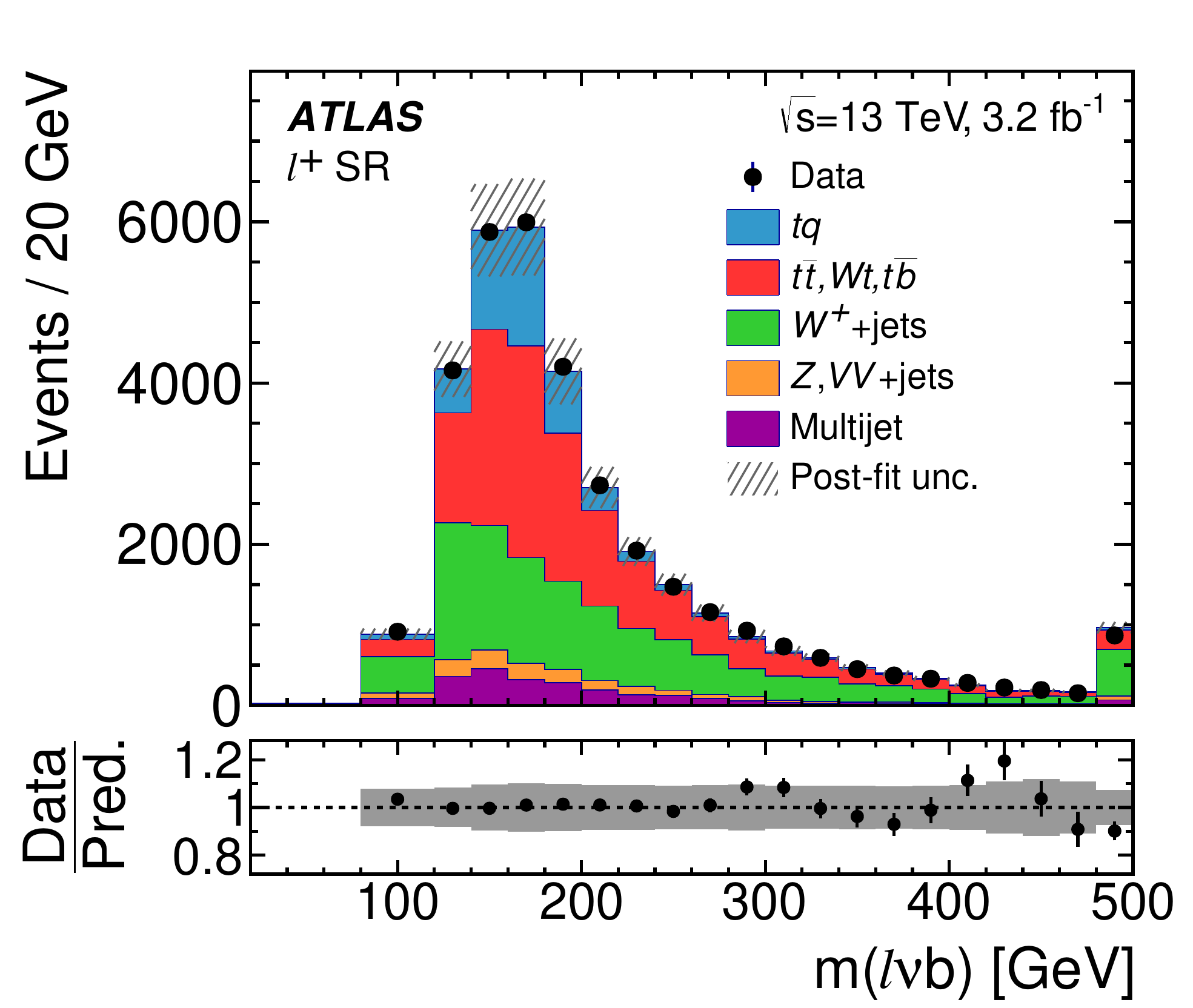}
\label{fig:postfit_mlnub_plus}
}
\subfigure[]{
\includegraphics[width=0.45\textwidth]{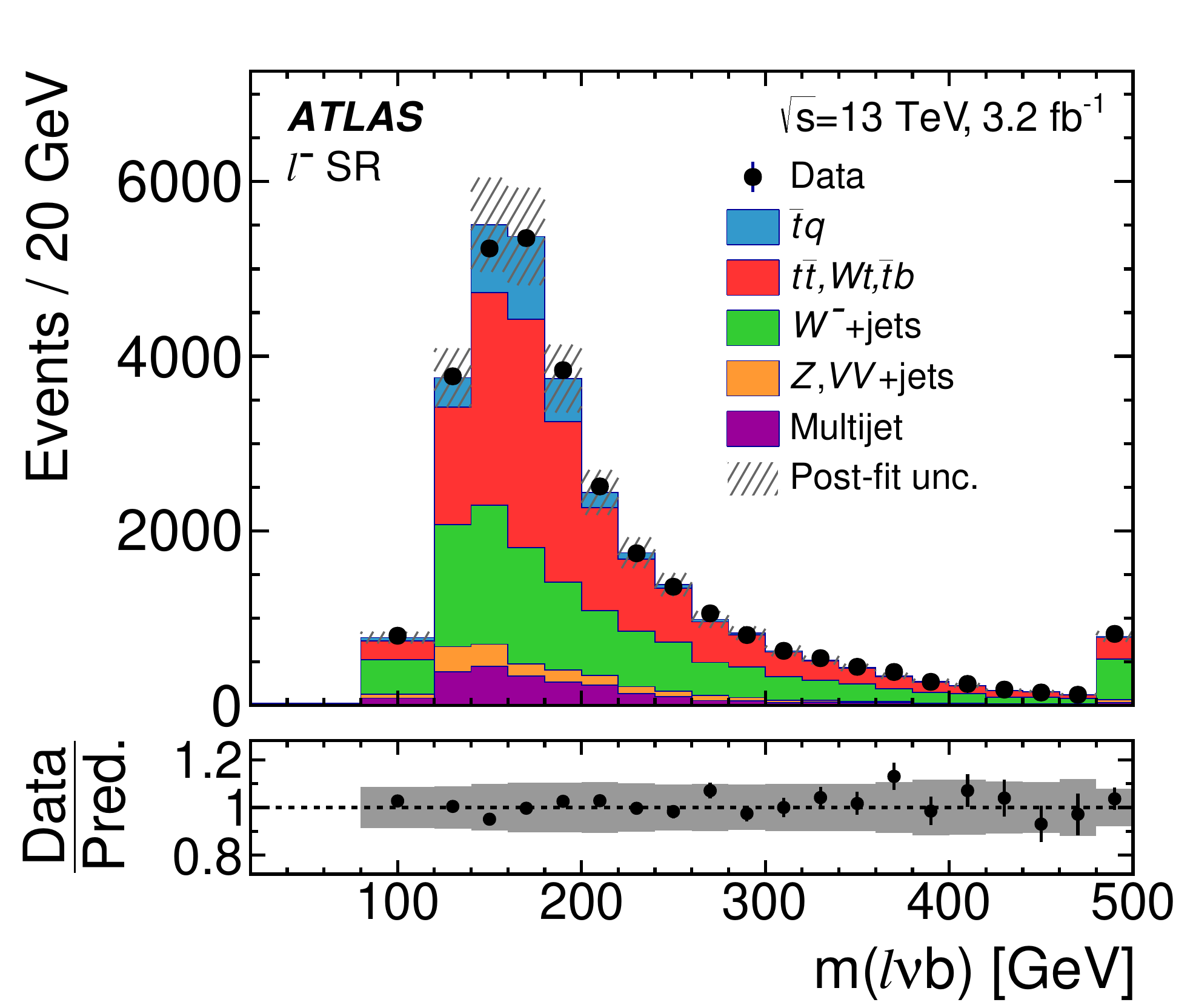}
\label{fig:postfit_mlnub_minus}
}
\subfigure[]{
\includegraphics[width=0.45\textwidth]{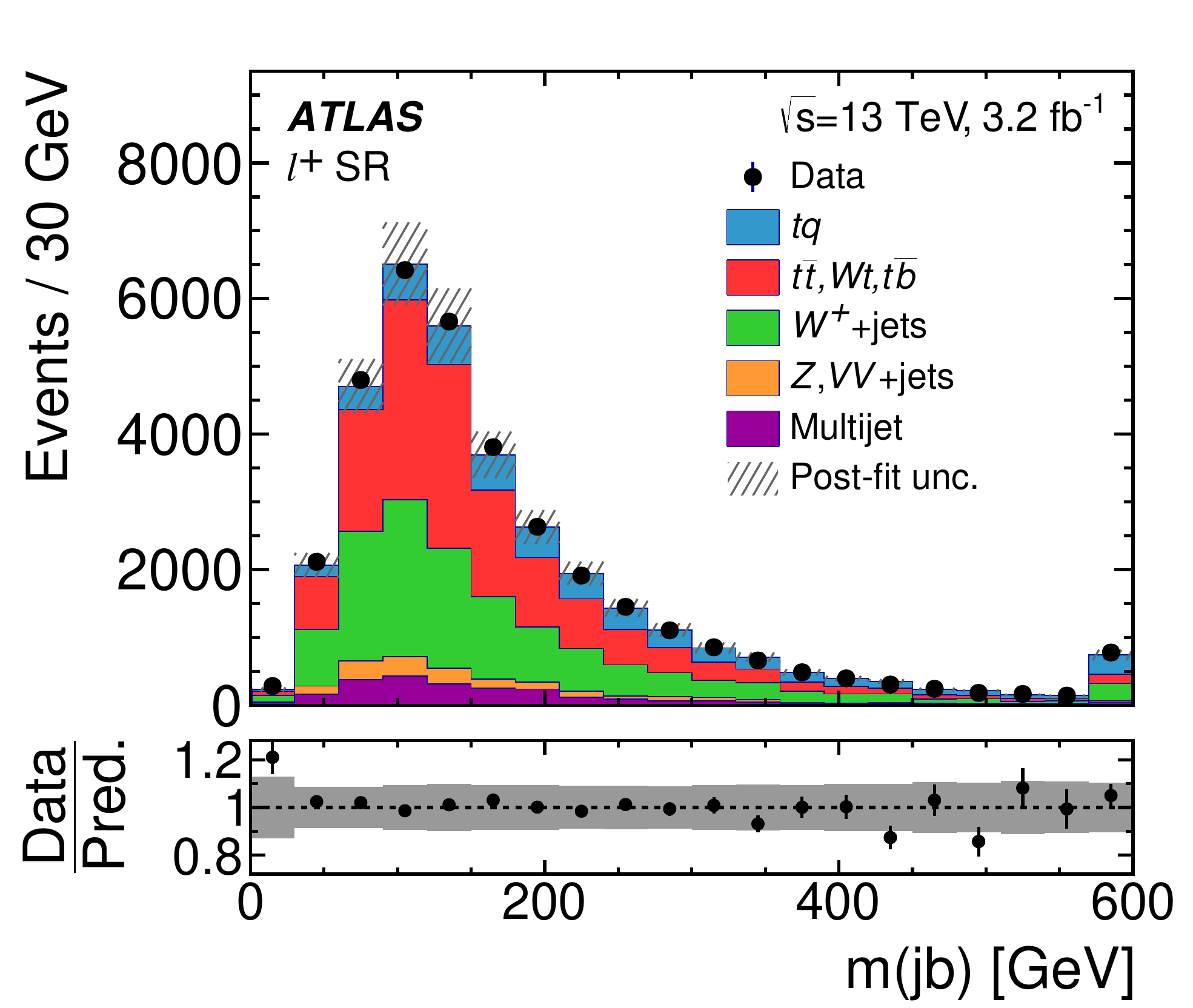}
\label{fig:postfit_mjj_plus}
}
\subfigure[]{
\includegraphics[width=0.45\textwidth]{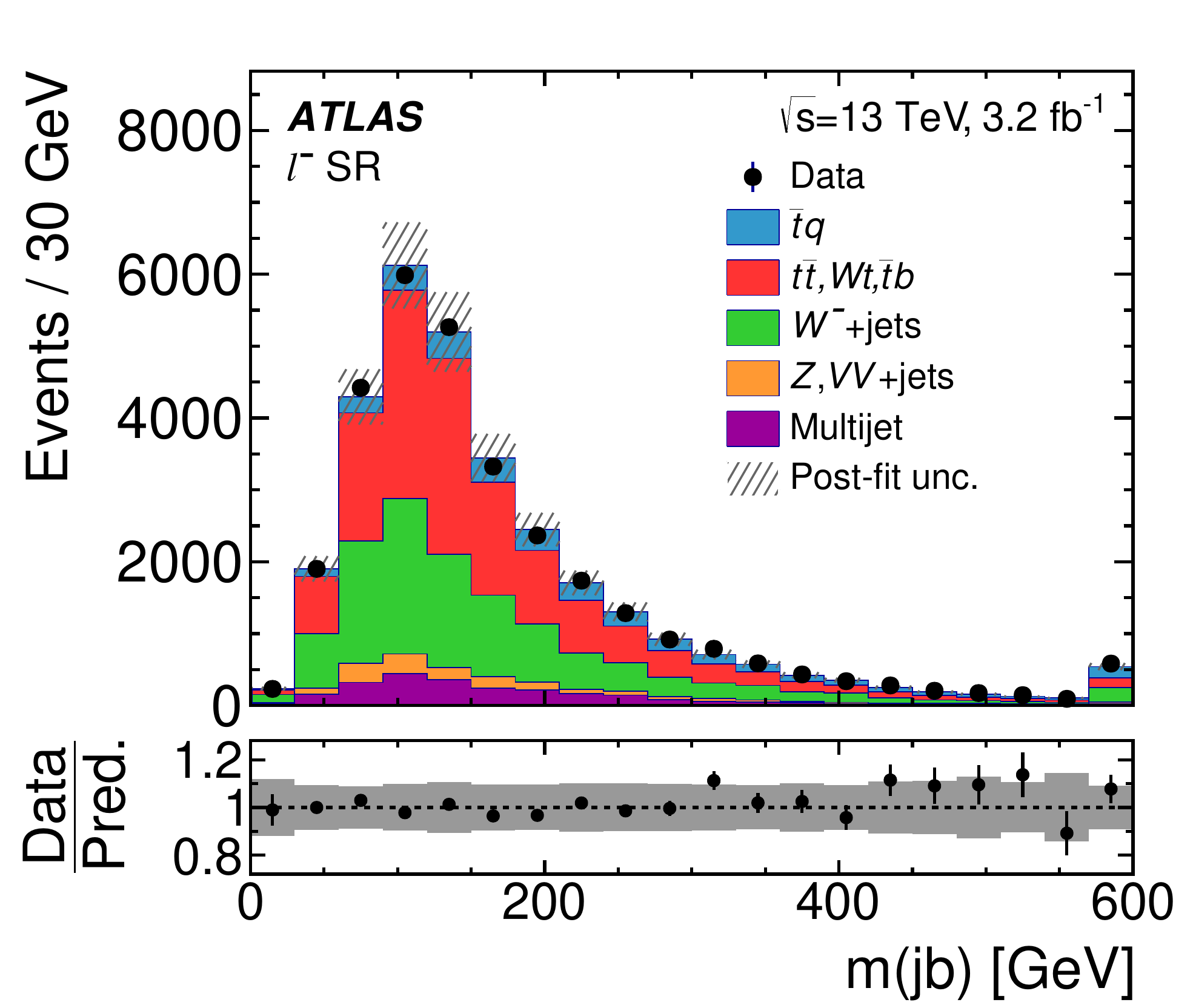}
\label{fig:postfit_mjj_minus}
}
\subfigure[]{
\includegraphics[width=0.45\textwidth]{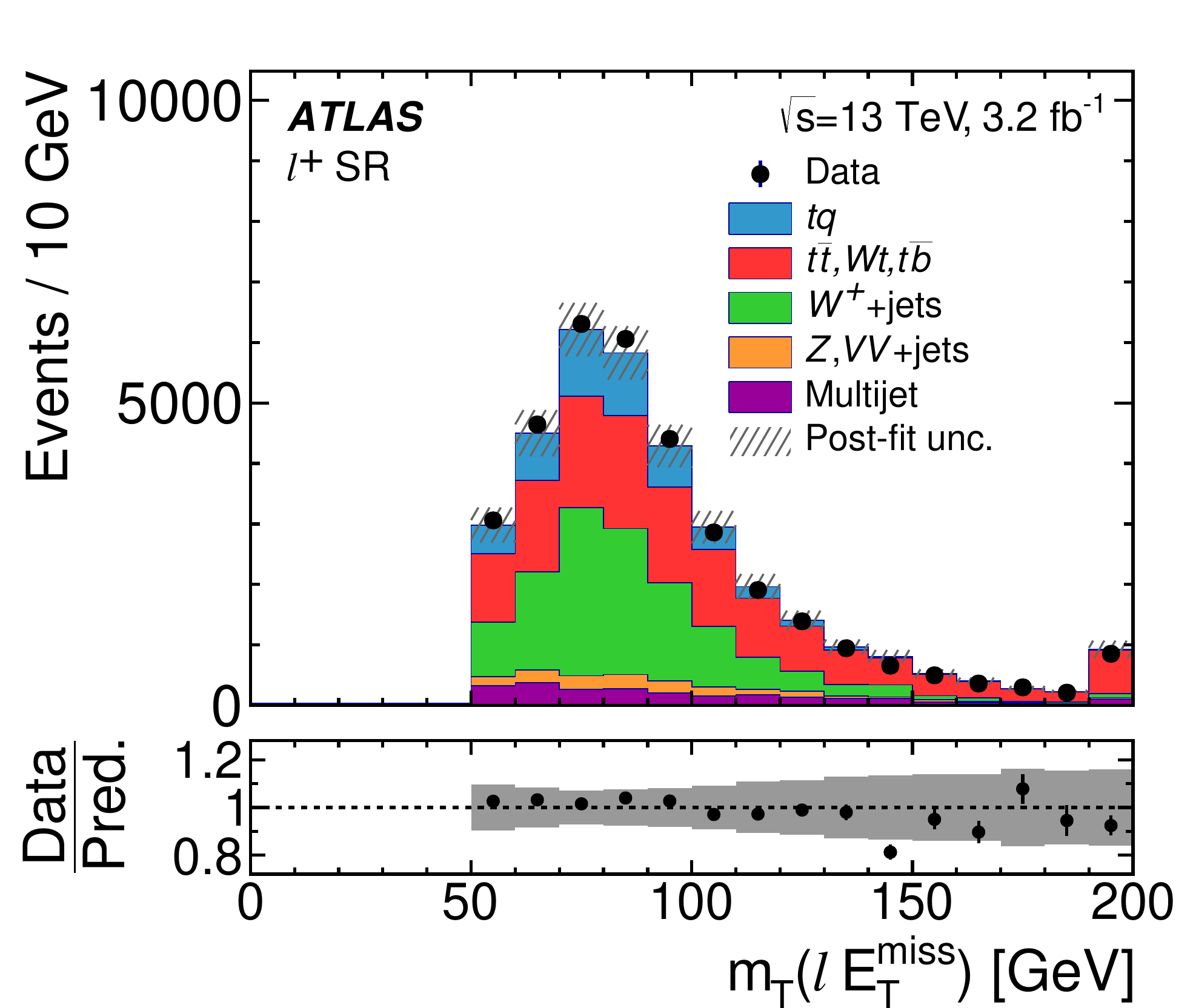}
\label{fig:postfit_mtw_plus}
}
\subfigure[]{
\includegraphics[width=0.45\textwidth]{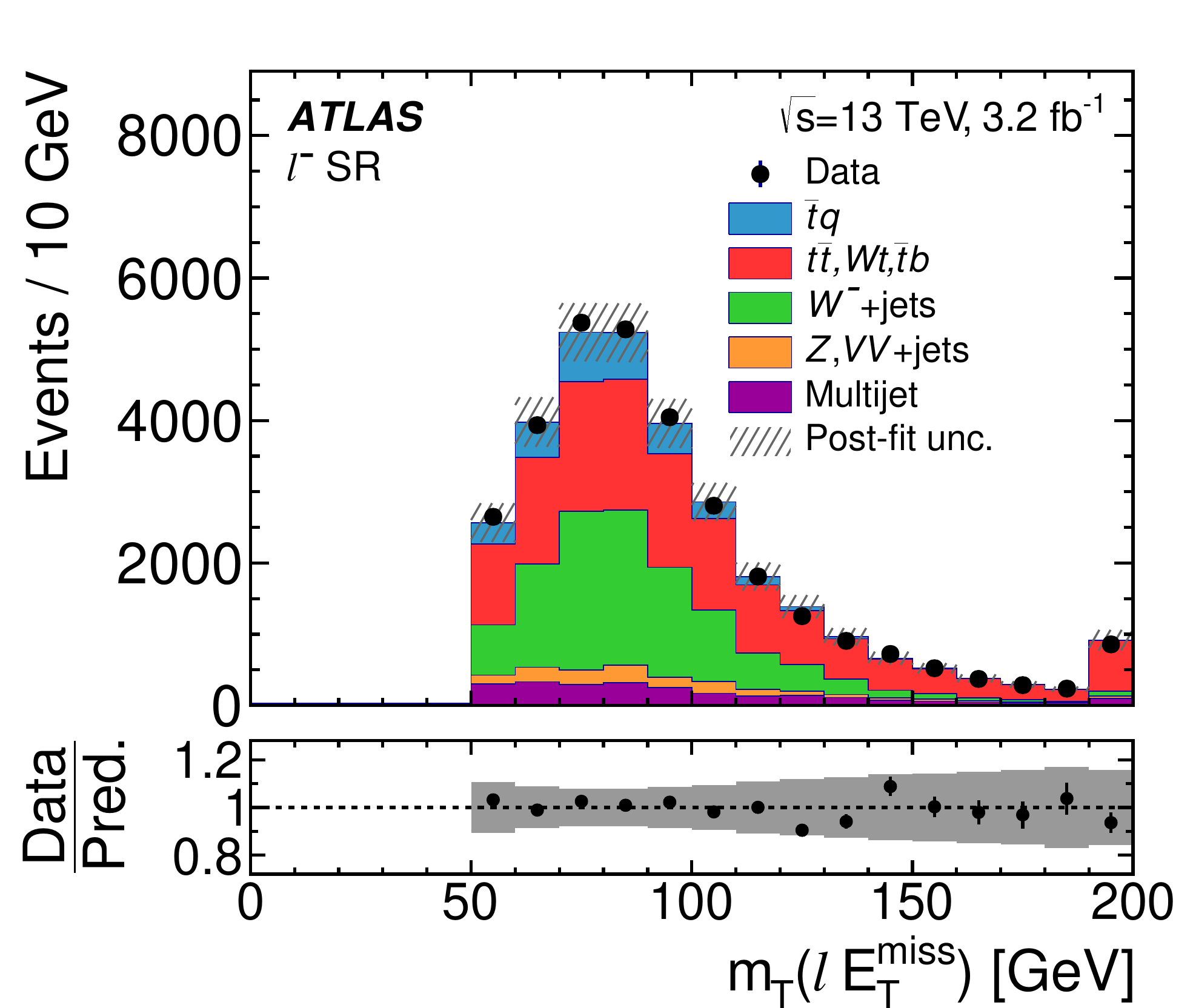}
\label{fig:postfit_mtw_minus}
}
\vspace{-0.5cm}
 \caption{Distributions of the three most important variables (left) for the \lpc  and (right) for the \lmc  
          normalised to the fit result.
          \subref{fig:postfit_mlnub_plus}-\subref{fig:postfit_mlnub_minus}: 
          reconstructed top-quark mass $m(\ell\nu b)$, 
          \subref{fig:postfit_mjj_plus}-\subref{fig:postfit_mjj_minus}: 
          invariant mass of the jet pair $m(j b)$, \subref{fig:postfit_mtw_plus}-\subref{fig:postfit_mtw_minus}:
          transverse mass of the $W$ boson \mtw.
          The hatched and grey error bands represents the post-fit uncertainty.
          The ratio of observed to predicted (Pred.) number of events in each bin is shown in the lower histogram.
          Events beyond the $x$-axis range are included in the last bin.}
 \label{fig:inputvar_NNFit}
\end{center}
\end{figure}

The fit finds the minimum of the negative log-likelihood 
function for the parameter values shown in Table~\ref{tab:betafit}.
Figure~\ref{fig:NNFit} compares the observed NN discriminant distributions to the
compound model of signal and backgrounds with each contribution normalised to the fit 
results from Table~\ref{tab:betafit}. 
The three most discriminating variables are presented in Figure~\ref{fig:inputvar_NNFit}. 
The model agrees with the data, within uncertainties.

\renewcommand\arraystretch{1.20}
\begin{table}[b]
\sisetup{round-mode=figures}
\begin{center}
  \begin{tabular}{lc
                   S[table-format=6.0,round-precision=3]@{$\,\pm\,$}S[table-format=3.0]
                   S[table-format=5.0,round-precision=3]@{$\,\pm\,$}S[table-format=3.0]}
\toprule
 Process          & $\hat{\beta}$   & \multicolumn{2}{c}{$\hat{\nu}$($\ell^{+}$)} & \multicolumn{2}{c}{$\hat{\nu}$($\ell^{-}$)}  \\
\midrule
 $tq$               & $1.15 \pm 0.03$  & 4835 & 142 &  \multicolumn{2}{c}{--}   \\
 $\bar{t}q$         & $1.12 \pm 0.05$  & \multicolumn{2}{c}{--}  & 3042 & 125    \\
\midrule
 $\ttbar,Wt,t\bar{b}+\bar{t}b$  & $0.91 \pm 0.03 $  & 13708 & 509 & 13644 & 507 \\
 $W^++$ jets           & $1.13 \pm 0.05 $  & 11994 & 552 & \multicolumn{2}{c}{--} \\
 $W^-+$ jets           & $1.21 \pm 0.06 $  & \multicolumn{2}{c}{--} & 10529 & 549 \\
 $Z,VV+$ jets          & --   & 
 \multicolumn{1}{S[table-format=5.0, round-precision=3]@{\phantom{$\,\pm\,$}}}{1531}
 & & 
 \multicolumn{1}{S[table-format=5.0, round-precision=3]@{\phantom{$\,\pm\,$}}}{1406} & \\
 Multijet background & --   &
 \multicolumn{1}{S[table-format=5.0, round-precision=3]@{\phantom{$\,\pm\,$}}}{2421}
 & & 
 \multicolumn{1}{S[table-format=5.0, round-precision=3]@{\phantom{$\,\pm\,$}}}{2421} & \\
 \midrule
 Total estimated & -- & 34489 & 764 & 31041 & 758 \\
 \midrule
 Total observed  & -- & \multicolumn{1}{S[table-format=5.0, round-mode=off]@{\phantom{$\,\pm\,$}}}{34459} &
                      & \multicolumn{1}{S[table-format=5.0, round-mode=off]@{\phantom{$\,\pm\,$}}}{31056} & \\
\bottomrule
\end{tabular}
\caption{\label{tab:betafit}Estimated scale factors, $\hat{\beta}$, and number of events, 
         $\hat{\nu}=\hat{\beta} \cdot \nu$, for the $\ell^{+}$ and $\ell^{-}$ channel from the minimisation
         of the likelihood function.  
         The quoted uncertainties in $\hat{\beta}$ and $\hat{\nu}$ include the statistical uncertainty and the
         uncertainties from the constraints on the background normalisation as used in the likelihood function.}
\end{center}
\end{table}

The systematic uncertainties in the cross-section measurements are determined
from pseudo-experiments which vary the signal acceptance, the background rates, 
and the shape of the NN discriminant.
By using samples of simulated events with variations reflecting the sources of systematic uncertainty, 
all of the effects are estimated and the pseudo-experiments are varied accordingly.
Rate and shape uncertainties are treated in a correlated way.
All systematic uncertainties apart from those related to the Monte Carlo statistics are also treated in a correlated way 
between the \lpc and the \lmc.
Table~\ref{tab:xs-uncertainty-obs} shows the 
contributions to the total uncertainty in the inclusive cross-section measurements. The table 
provides the uncertainties estimated for the observed signal and background
rates as obtained from the maximum-likelihood fit to the observed
collision data. Uncertainties in the extrapolation to the full phase space are included 
in the generator-related uncertainties. 
\begin{table*}[p]
\begin{center}
\begin{tabular}{lccc}
    \toprule
    Source &  $\displaystyle \frac{\Delta \sigma(tq)}{\sigma(tq)}$ [\%]   
           &  $\displaystyle \frac{\Delta \sigma(\bar{t}q)}{\sigma(\bar{t}q)} \ [\%] $ 
           &  $\displaystyle \frac{\Delta R_{t}}{R_{t}} \ [\%]$  \\
    \midrule
    Data statistics         & $\pm$ 2.9 & $\pm$ 4.1  & $\pm$ 5.0 \\
    Monte Carlo statistics    & $\pm$ 2.8 & $\pm$ 4.2 & $\pm$ 5.1 \\
    \midrule
    \multicolumn{4}{c}{\textbf{Reconstruction efficiency and calibration uncertainties}} \\
    Muon uncertainties         & $\pm$ 0.8 & $\pm$ 0.9 & $\pm$ 1.0\\
    Electron uncertainties     & $< 0.5$ & $\pm$ 0.5 & $\pm$ 0.7\\
    JES                        & $\pm$ 3.4 & $\pm$ 4.1 & $\pm$ 1.2\\
    Jet energy resolution      & $\pm$ 3.9 & $\pm$ 3.1 & $\pm$ 1.1\\
    $\MET$ modelling           & $\pm$ 0.9 & $\pm$ 1.2 & $< 0.5$\\
    $b$-tagging efficiency     & $\pm$ 7.0 & $\pm$ 6.9 & $< 0.5$\\
    $c$-tagging efficiency     & $< 0.5$ & $\pm$ 0.5 & $\pm$ 0.6 \\
    Light-jet tagging efficiency & $< 0.5$ & $< 0.5$ & $< 0.5$ \\ 
    Pile-up reweighting       & $\pm$ 1.5 & $\pm$ 2.2 & $\pm$ 3.8\\
    \multicolumn{4}{c}{\textbf{Monte Carlo generators}} \\
    $tq$ parton shower generator      & $\pm$ 13.0 & $\pm$ 14.3 & $\pm$ 1.9\\
    $tq$ NLO matching          & $\pm$ 2.1 & $\pm$ 0.7 & $\pm$ 2.8\\
    $tq$ radiation             & $\pm$ 3.7 & $\pm$ 3.4 & $\pm$ 3.7\\ 
    $\ttbar$, $Wt$, $t\bar{b}+\bar{t}b$ parton shower generator  & $\pm$ 3.2 & $\pm$ 4.4& $\pm$ 1.2\\
    $\ttbar$, $Wt$, $t\bar{b}+\bar{t}b$ NLO matching  & $\pm$ 4.4 & $\pm$ 8.6& $\pm$ 4.6\\    
    $\ttbar$, $Wt$, $t\bar{b}+\bar{t}b$ radiation     & $< 0.5$ & $\pm$ 1.1& $\pm$ 0.7\\
    \multicolumn{3}{c}{\vspace*{-3mm}} \\
    PDF                        & $\pm$ 0.6 & $\pm$ 0.9& $< 0.5$\\
    \multicolumn{4}{c}{\textbf{Background normalisation}} \\
    Multijet normalisation   & $\pm$ 0.3 & $\pm$ 2.0 & $\pm$ 1.8\\
    Other background normalisation & $\pm$ 0.4 &  $\pm$ 0.5 & $< 0.5$ \\
    \multicolumn{3}{c}{\vspace*{-3mm}} \\
    Luminosity                 & $\pm$ 2.1 & $\pm$ 2.1 & $< 0.5$ \\
    \midrule
    Total systematic uncertainty          & $\pm$ 17.5 & $\pm$ 20.0& $\pm$ 10.2\\ 
    Total uncertainty                     & $\pm$ 17.8 & $\pm$ 20.4& $\pm$ 11.4\\
    \bottomrule
   \end{tabular}
   \caption{\label{tab:xs-uncertainty-obs}
   List of systematic uncertainties contributing to the total uncertainty
   in the measured values of $\sigma(tq)$, $\sigma(\bar{t}q)$, 
   and $R_t= \sigma(tq)/\sigma(\bar{t}q)$.
   The estimation of the systematic uncertainties has a statistical uncertainty of \SI{0.3}{\%}. 
   Uncertainties contributing less than \SI{0.5}{\%} are marked with ``$<0.5$''
   }
\end{center}
\end{table*}

\clearpage

\section{Cross-section measurement}

\label{sec:result}

\begin{figure*}[!ht]
  \centering
  \includegraphics[width=0.80\textwidth]{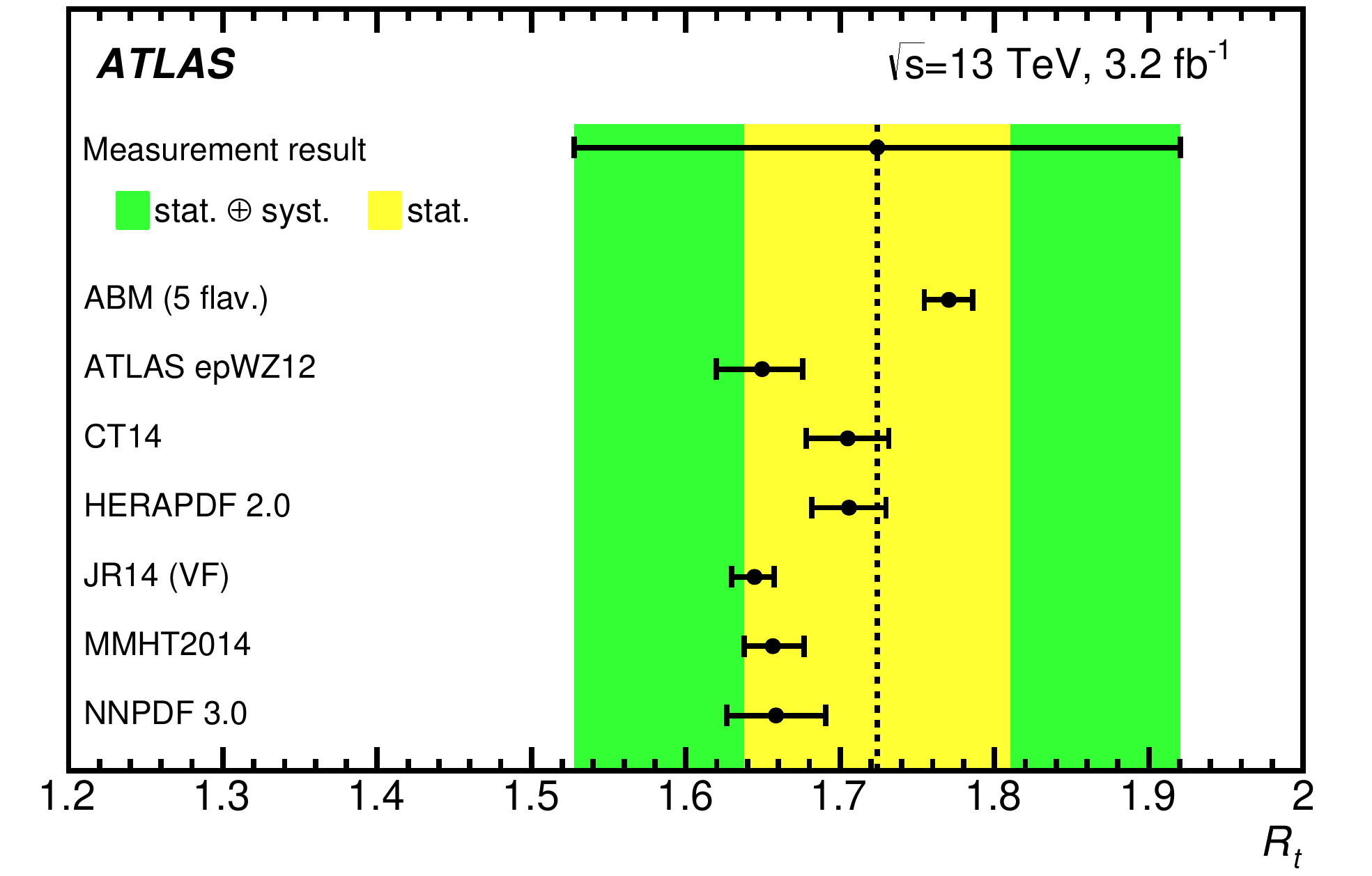}
  \caption{\label{fig:rtop} Comparison between observed and predicted values of $R_t=\frac{\sigma_{t}}{\sigma_{\bar{t}}}$. 
  Predictions are calculated at NLO precision~\cite{Campbell:2009ss,Kant:2014oha} in the five-flavour scheme and 
  given for different NLO PDF sets. The uncertainty includes the uncertainty in the renormalisation and
  factorisation scales, as well as the combined internal PDF and \alphas uncertainty. 
  The dotted black line indicates the measured value. 
  The combined statistical and systematic uncertainty of the measurement is shown in green, 
  while the statistical uncertainty is represented by the yellow error band.
  Predictions for all presented PDF sets are within the statistical uncertainty of the measurement.
  }
\end{figure*}

After performing the binned maximum-likelihood fit to the NN discriminant distribution 
and estimating the total uncertainty, the inclusive cross-sections 
of top-quark and top-antiquark production in the $t$-channel are measured to be:
\begin{eqnarray*}
\sigma(tq)      &=&  156 \pm 5  \, (\mathrm{stat.}) \pm 27 \, (\mathrm{syst.}) \pm 3\,(\mathrm{lumi.})~\text{pb}\\
\sigma(\tbar q) &=&   91 \pm 4  \, (\mathrm{stat.}) \pm 18 \, (\mathrm{syst.}) \pm 2\,(\mathrm{lumi.})~\text{pb}\\
R_t             &=&  1.72 \pm 0.09  \, (\mathrm{stat.}) \pm 0.18 (\mathrm{syst.}),\\
\end{eqnarray*}
assuming a top-quark mass of $\mtop = \SI{172.5}{\GeV}$.
Figure~\ref{fig:rtop} compares the measured value of $R_t$ to NLO predictions~\cite{Campbell:2009ss}
obtained with Hathor~\cite{Kant:2014oha} using different PDF sets.
PDF sets are available from various groups worldwide: \textsc{CTEQ}~\cite{Lai:2010vv,Dulat:2015mca},
\mstw~\cite{Martin:2009iq}/\mmht~\cite{Harland-Lang:2014zoa}, \nnpdf~\cite{Ball:2011mu,Ball:2014uwa}, 
\textsc{JR}~\cite{Jimenez-Delgado:2014twa}, \textsc{ABM}~\cite{Alekhin:2012ig}, and
\hera~\cite{Aaron:2009wt,Abramowicz:2015mha}. Also, the first PDF
set provided by the ATLAS Collaboration is considered~\cite{STDM-2011-43}.
The PDFs provided by the different groups differ in the data used, the value of \alphas, the values of the quark
masses, and the treatment of heavy-quark masses. Other differences concern the way higher-order corrections are implemented, 
the parametrisation of the PDF fitting-model, the way of treating systematic uncertainties and the criteria for estimating confidence levels.
Uncertainties in the predicted values include the uncertainty in the renormalisation and factorisation scales 
and the combined PDF and \alphas uncertainty of the respective PDF set.
All PDF predictions are in agreement with the measurement.  

For the purpose of determining $\fl \cdot |V_{tb}|$, the measured inclusive cross-sections of the $tq$ and the
$\tbar q$ process are combined, assuming that each uncertainty is \SI{100}{\%}
correlated between the two channels. The statistical
uncertainty of the data and the uncertainty due to the limited size of the MC samples, 
are treated as uncorrelated.
The combined cross-section is calculated to be:
\begin{eqnarray*}
\sigma(tq+\tbar q) &=&  247 \pm 6  \, (\mathrm{stat.}) \pm 45 \, (\mathrm{syst.}) \pm 5\,(\mathrm{lumi.})~\text{pb}\\
                   &=&  247 \pm 46~\text{pb}.
\end{eqnarray*}

To estimate the dependence of the measured cross-sections on the assumed top-quark mass,
simulated samples with $\mtop = \SI{170}{\GeV}$ and $\mtop = \SI{175}{\GeV}$ are used.
The measurement is repeated for each top-quark mass.
In Table~\ref{tab:paratopmass_app}, the measured cross-sections and their ratio are given.
\begin{table}[!ht]
\begin{center}
   \begin{tabular}{c|c|c|c|c }
    \toprule
	$\mtop$ [\si{\GeV}] & $\sigma(tq)$ [pb] & $\sigma(\tbar q)$ [pb] & $\sigma(tq+\tbar q)$ [pb] & $R_t$ \\
    \midrule 
    170.0 & $156\pm 5$ & $93\pm 4$ & $249\pm 6$ &  $1.69\pm 0.09$ \\
    172.5 & $156\pm 5$ & $91\pm 4$ & $247\pm 6$ &  $1.72\pm 0.09$ \\
	175.0 & $155\pm 5$ & $92\pm 4$ & $247\pm 6$ &  $1.68\pm 0.09$ \\
    \bottomrule
   \end{tabular}
\caption{Measured values of the cross-sections $\sigma(tq)$, $\sigma(\tbar
q)$, $\sigma_{\mathrm{tot}}(tq+\tbar q)$, and $R_t$ for different simulated
top-quark masses.
The quoted uncertainties are statistical only.}
\label{tab:paratopmass_app}
\end{center}
\end{table}

Single top-quark production in the $t$-channel proceeds via a $Wtb$ 
vertex and the measured cross-section is proportional to $(\fl \cdot |V_{tb}|)^2$ as discussed in
Section~\ref{sec:intro}.
The $\fl \cdot |V_{tb}|$ measurement via single top-quark production is independent of assumptions 
about the number of quark generations or about the unitarity of the CKM matrix.
The assumptions made are: $|V_{tb}|$ is much bigger than $|V_{td}|$ and $|V_{ts}|$,
which is in agreement with the measurement of
$R={\cal B}(t \rightarrow Wb)/\sum_{q=d,s,b} {\cal B}(t\rightarrow Wq)$~\cite{CMS-TOP-12-035},
the top quark decays exclusively into a $W$ boson and a $b$-quark, and the
$Wtb$ interaction is a left-handed weak coupling like that in the SM.
A strategy to relax the first two assumptions and account for production and decay of top quarks
via $Wts$ and $Wtd$ vertices is delineated in \Ref{\cite{Lacker:2012ek}}.

The value of $\fl \cdot |V_{tb}|$ is extracted by dividing the measured $\sigma(tq+\tbar q) = 247 \pm 46~\text{pb}$
by its value predicted at NLO, $\sigma^{\text{th}}(tq+\tbar q) = 217 \pm 10~\text{pb}$. 
Changes in $\fl \cdot |V_{tb}|$ also affect $Wt$ and $t\bar{b}+\bar{t}b$ production. However, their
contributions are small and their variation does not change the $t$-channel fit result.
The result obtained is 
\begin{eqnarray*}
 \label{eq:Vtb}
\fl \cdot |V_{tb}| & = & 1.07 \pm 0.01 \, (\mathrm{stat.}) \pm 0.09 \, (\mathrm{syst.}) \pm 0.02 \,
(\mathrm{theor.}) \pm 0.01 \, (\mathrm{lumi.}) \\
 & = & 1.07 \pm 0.09. \nonumber
\end{eqnarray*}
The experimental uncertainty is 0.09, including the statistical uncertainty, 
the systematic uncertainties, and the uncertainty in the luminosity.
The theoretical uncertainty is 0.02, including scale uncertainties and PDF uncertainties.

Setting \fl=1 as required by the SM, and assuming a uniform prior of 
one in $|V_{tb}|^2$ in the interval $[0,1]$ and a Gaussian-shaped 
likelihood curve for  $|V_{tb}|^2$, 
a Bayesian lower limit giving $|V_{tb}|>0.84$ at \SI{95}{\%} CL, is obtained.

\FloatBarrier

\section{Conclusion} 
\label{sec:conclusion}
A measurement of the $t$-channel single-top-quark and single-top-antiquark production cross-sections is performed in
events with a leptonically decaying $W$ boson with \SI{3.2}{\per\fb} of $pp$ collision data at $\sqrt{s} =
\SI{13}{\TeV}$ recorded with the ATLAS detector at the LHC in 2015.
Events are selected by requiring exactly one electron or muon, missing transverse momentum, 
and two jets with high transverse momentum, exactly one of which is required to be $b$-tagged.

A binned maximum-likelihood fit to neural-network discriminant distributions yields the following cross-sections:
\begin{eqnarray*}
\sigma(tq) &=& 156 \pm 28 ~\text{pb},\\
\sigma(\tbar q) &=& 91 \pm 19 ~\text{pb},\\
\sigma(tq+\tbar q)  &=&  247 \pm 46~\text{pb} \\
\end{eqnarray*}
in agreement with SM predictions.
The cross-section ratio of $tq$ and $\tbar q$ production is found to be $R_t = 1.72 \pm 0.20$.
The coupling at the $Wtb$ vertex is determined to be $\fl \cdot |V_{tb}|=1.07 \pm 0.09$ and
a lower limit on the CKM matrix element is set, giving $|V_{tb}|$ > 0.84 at the
\SI{95}{\%} CL.
These measurements are dominated by systematic uncertainties,
from which the uncertainties connected with MC generators are the biggest ones.
Further improvements in these generators could lead to smaller 
expected uncertainties and therefore higher precision in the course of Run 2.

\section*{Acknowledgements}


We thank CERN for the very successful operation of the LHC, as well as the
support staff from our institutions without whom ATLAS could not be
operated efficiently.

We acknowledge the support of ANPCyT, Argentina; YerPhI, Armenia; ARC, Australia; BMWFW and FWF, Austria; ANAS, Azerbaijan; SSTC, Belarus; CNPq and FAPESP, Brazil; NSERC, NRC and CFI, Canada; CERN; CONICYT, Chile; CAS, MOST and NSFC, China; COLCIENCIAS, Colombia; MSMT CR, MPO CR and VSC CR, Czech Republic; DNRF and DNSRC, Denmark; IN2P3-CNRS, CEA-DSM/IRFU, France; GNSF, Georgia; BMBF, HGF, and MPG, Germany; GSRT, Greece; RGC, Hong Kong SAR, China; ISF, I-CORE and Benoziyo Center, Israel; INFN, Italy; MEXT and JSPS, Japan; CNRST, Morocco; FOM and NWO, Netherlands; RCN, Norway; MNiSW and NCN, Poland; FCT, Portugal; MNE/IFA, Romania; MES of Russia and NRC KI, Russian Federation; JINR; MESTD, Serbia; MSSR, Slovakia; ARRS and MIZ\v{S}, Slovenia; DST/NRF, South Africa; MINECO, Spain; SRC and Wallenberg Foundation, Sweden; SERI, SNSF and Cantons of Bern and Geneva, Switzerland; MOST, Taiwan; TAEK, Turkey; STFC, United Kingdom; DOE and NSF, United States of America. In addition, individual groups and members have received support from BCKDF, the Canada Council, CANARIE, CRC, Compute Canada, FQRNT, and the Ontario Innovation Trust, Canada; EPLANET, ERC, ERDF, FP7, Horizon 2020 and Marie Sk{\l}odowska-Curie Actions, European Union; Investissements d'Avenir Labex and Idex, ANR, R{\'e}gion Auvergne and Fondation Partager le Savoir, France; DFG and AvH Foundation, Germany; Herakleitos, Thales and Aristeia programmes co-financed by EU-ESF and the Greek NSRF; BSF, GIF and Minerva, Israel; BRF, Norway; CERCA Programme Generalitat de Catalunya, Generalitat Valenciana, Spain; the Royal Society and Leverhulme Trust, United Kingdom.

The crucial computing support from all WLCG partners is acknowledged gratefully, in particular from CERN, the ATLAS Tier-1 facilities at TRIUMF (Canada), NDGF (Denmark, Norway, Sweden), CC-IN2P3 (France), KIT/GridKA (Germany), INFN-CNAF (Italy), NL-T1 (Netherlands), PIC (Spain), ASGC (Taiwan), RAL (UK) and BNL (USA), the Tier-2 facilities worldwide and large non-WLCG resource providers. Major contributors of computing resources are listed in Ref.~\cite{ATL-GEN-PUB-2016-002}.

\printbibliography
\newpage 
\begin{flushleft}
{\Large The ATLAS Collaboration}

\bigskip

M.~Aaboud$^\textrm{\scriptsize 136d}$,
G.~Aad$^\textrm{\scriptsize 87}$,
B.~Abbott$^\textrm{\scriptsize 114}$,
J.~Abdallah$^\textrm{\scriptsize 8}$,
O.~Abdinov$^\textrm{\scriptsize 12}$,
B.~Abeloos$^\textrm{\scriptsize 118}$,
R.~Aben$^\textrm{\scriptsize 108}$,
O.S.~AbouZeid$^\textrm{\scriptsize 138}$,
N.L.~Abraham$^\textrm{\scriptsize 152}$,
H.~Abramowicz$^\textrm{\scriptsize 156}$
% ...
\bigskip
\\
% ...
$^{8}$ Department of Physics, The University of Texas at Arlington, Arlington TX, United States of America\\
% ...
$^{12}$ Institute of Physics, Azerbaijan Academy of Sciences, Baku, Azerbaijan\\
% ...
$^{87}$ CPPM, Aix-Marseille Universit{\'e} and CNRS/IN2P3, Marseille, France\\
% ...
$^{108}$ Nikhef National Institute for Subatomic Physics and University of Amsterdam, Amsterdam, Netherlands\\
% ...
$^{114}$ Homer L. Dodge Department of Physics and Astronomy, University of Oklahoma, Norman OK, United States of America\\
% ...
$^{118}$ LAL, Univ. Paris-Sud, CNRS/IN2P3, Universit{\'e} Paris-Saclay, Orsay, France\\
% ...
$^{136}$ $^{(a)}$ Facult{\'e} des Sciences Ain Chock, R{\'e}seau Universitaire de Physique des Hautes Energies - Universit{\'e} Hassan II, Casablanca; $^{(b)}$ Centre National de l'Energie des Sciences Techniques Nucleaires, Rabat; $^{(c)}$ Facult{\'e} des Sciences Semlalia, Universit{\'e} Cadi Ayyad, LPHEA-Marrakech; $^{(d)}$ Facult{\'e} des Sciences, Universit{\'e} Mohamed Premier and LPTPM, Oujda; $^{(e)}$ Facult{\'e} des sciences, Universit{\'e} Mohammed V, Rabat, Morocco\\
% ...
$^{138}$ Santa Cruz Institute for Particle Physics, University of California Santa Cruz, Santa Cruz CA, United States of America\\
% ...
$^{152}$ Department of Physics and Astronomy, University of Sussex, Brighton, United Kingdom\\
% ...
$^{156}$ Raymond and Beverly Sackler School of Physics and Astronomy, Tel Aviv University, Tel Aviv, Israel\\
% ...
$^{a}$ Also at Department of Physics, King's College London, London, United Kingdom\\
$^{b}$ Also at Institute of Physics, Azerbaijan Academy of Sciences, Baku, Azerbaijan\\
$^{c}$ Also at Novosibirsk State University, Novosibirsk, Russia\\
$^{d}$ Also at TRIUMF, Vancouver BC, Canada\\
$^{e}$ Also at Department of Physics {\&} Astronomy, University of Louisville, Louisville, KY, United States of America\\
% ...
\end{flushleft}


\end{document}